\documentclass[pra, twocolumn,amsfonts,amsmath,amssymb,eufrak]{revtex4-1}

\usepackage{epsfig}
\usepackage{color}
\usepackage{bm}
\usepackage{hyperref}
\usepackage[dvipsnames]{xcolor}
\hypersetup{
    bookmarksnumbered=true, 
    unicode=false, 
    pdfstartview={FitH}, 
    pdftitle={}, 
    pdfauthor={}, 
    pdfsubject={}, 
    pdfcreator={}, 
    pdfproducer={}, 
    pdfkeywords={}, 
    pdfnewwindow=true, 
    colorlinks=true, 
    linkcolor=NavyBlue, 
    citecolor=NavyBlue, 
    filecolor=NavyBlue, 
    urlcolor=NavyBlue 
}

\usepackage{braket}

\usepackage{comment}

\usepackage{graphicx}
\usepackage{amsthm}

\usepackage{ragged2e}

\newcommand{\eq}[1]{\begin{equation} #1 \end{equation}}
\newcommand{\eqa}[2]{\begin{equation} #1 \label{#2} \end{equation}}
\newcommand{\balign}[1]{\begin{align} #1 \end{align}}

\newcommand{\ul}{\underline}

\newcommand{\figin}[4]
{\begin{figure}[tb]
\includegraphics[width= #1]{#2.pdf}
\caption{#3}
\label{f:#4}
\end{figure}}

\newcommand{\todayd}{\the\year/\the\month/\the\day}

\newcommand{\bib}{\bibitem}

\newcommand{\Lmd}{\Lambda}

\newcommand{\lb}{\label}
\newcommand{\nt}{\notag}
\newcommand{\Tr}{\mathrm{Tr}}

\newcommand{\bel}{\begin{easylist}}
\newcommand{\eel}{\end{easylist}}
\newcommand{\bi}[1]{\begin{itemize} #1 \end{itemize}}
\newcommand{\be}[1]{\begin{enumerate} #1 \end{enumerate}}

\newcommand{\eref}[1]{Eq.~\eqref{#1}}
\newcommand{\tref}[1]{Theorem~\ref{t:#1}}
\newcommand{\dref}[1]{Definition~\ref{t:#1}}
\newcommand{\lref}[1]{Lemma~\ref{t:#1}}
\newcommand{\pref}[1]{Proposition~\ref{t:#1}}
\newcommand{\fref}[1]{Fig.~\ref{f:#1}}
\newcommand{\sref}[1]{Sec.~\ref{s:#1}}

\def \({\left(}
\def \){\right)}
\def \[{\left[}
\def \]{\right]}

\newcommand{\abs}[1]{\left|#1\right|}

\newcommand{\sumtwo}[2]%
{\mathop{\sum_{#1}}_{#2}}
\newcommand{\sumthree}[3]%
{\mathop{\mathop{\sum_{#1}}_{#2}}_{#3}}
\newcommand{\sumfour}[4]%
{\mathop{\mathop{\mathop{\sum_{#1}}_{#2}}_{#3}}_{#4}} 
\newcommand{\prodtwo}[2]%
{\mathop{\prod_{#1}}_{#2}}
\newcommand{\mintwo}[2]%
{\mathop{\min_{#1}}_{#2}}
\newcommand{\maxtwo}[2]%
{\mathop{\max_{#1}}_{#2}}
\newcommand{\maxthree}[3]%
{\mathop{\mathop{\max_{#1}}_{#2}}_{#3}}
\newcommand{\limtwo}[2]%
{\mathop{\lim_{#1}}_{#2}}
\newcommand{\suptwo}[2]%
{\mathop{\sup_{#1}}_{#2}}
\newcommand{\supthree}[3]%
{\mathop{\mathop{\sup_{#1}}_{#2}}_{#3}}
\newcommand{\supfour}[4]%
{\mathop{\mathop{\mathop{\sup_{#1}}_{#2}}_{#3}}_{#4}} 
\newcommand{\inftwo}[2]%
{\mathop{\inf_{#1}}_{#2}}
\newcommand{\infthree}[3]%
{\mathop{\mathop{\inf_{#1}}_{#2}}_{#3}}
\newcommand{\inffour}[4]%
{\mathop{\mathop{\mathop{\inf_{#1}}_{#2}}_{#3}}_{#4}} 

\newcommand\calE{{\cal E}}

\newcommand\calM{{\cal M}}
\newcommand\calN{{\cal N}}

\newcommand\calT{{\cal T}}

\newcommand{\bsp}{\boldsymbol{p}}
\newcommand{\bsq}{\boldsymbol{q}}

\newcommand{\bbN}{\mathbb{N}}
\newcommand{\bbO}{\mathbb{O}}
\newcommand{\bbQ}{\mathbb{Q}}
\newcommand{\bbR}{\mathbb{R}}

\newcommand{\ep}{\varepsilon}

\newcommand{\bcs}{\backslash}
\newcommand{\Di}{\mathit{\Delta}}

\newcommand{\para}[1]{{\em #1}\/.---}

\newtheorem{thm}{Theorem}
\newtheorem{lm}[thm]{Lemma}
\newtheorem{pro}[thm]{Proposition}

\newcommand{\bthm}[1]{\begin{thm} #1 \end{thm}}
\newcommand{\blm}[1]{\begin{lm} #1 \end{lm}}
\newcommand{\bpro}[1]{\begin{pro} #1 \end{pro}}

\theoremstyle{definition}
\newtheorem{dfn}{Definition}
\newcommand{\bdf}[1]{\begin{dfn} #1 \end{dfn}}

\newcommand{\rhoG}{\rho_{\rm Gibbs}}
\newcommand{\rhopG}{\rho'_{\rm Gibbs}}
\newcommand{\test}{t_{\rm est}}

\newcommand{\pG}{\bsp_{\rm Gibbs}}

\def\rnum#1{\resizebox{0.5em}{\height}{\expandafter{\romannumeral #1}}}
\def\Rnum#1{\resizebox{0.5em}{\height}{\uppercase\expandafter{\romannumeral #1}}}

\makeatletter
\def\shorttableofcontents#1#2{\bgroup\c@tocdepth=#2\@restonecolfalse
  \settowidth\js@tocl@width{\headfont\prechaptername\postchaptername}%
  \settowidth\@tempdima{\headfont\appendixname}%
  \ifdim\js@tocl@width<\@tempdima \setlength\js@tocl@width{\@tempdima}\fi
  \ifdim\js@tocl@width<2zw \divide\js@tocl@width by 2 \advance\js@tocl@width 1zw\fi
\if@tightshtoc
    \parsep\z@
  \fi
  \if@twocolumn\@restonecoltrue\onecolumn\fi
  \@ifundefined{chapter}%
  {\section*{{#1}
        \@mkboth{\uppercase{#1}}{\uppercase{#1}}}}%
  {\chapter*{{#1}
        \@mkboth{\uppercase{#1}}{\uppercase{#1}}}}%
  \@startshorttoc{toc}\if@restonecol\twocolumn\fi\egroup}
\makeatother

\begin{document}


\title{
Quantum thermodynamics with coherence: Covariant Gibbs-preserving operation is characterized by the free energy
}

\author{Naoto Shiraishi} 
\email{shiraishi@phys.c.u-tokyo.ac.jp}
\affiliation{Department of Basic Science, The University of Tokyo, 3-8-1 Komaba, Meguro-ku, Tokyo 153-8902, Japan}%



\begin{abstract}
The resource theory with covariant Gibbs-preserving operations, also called enhanced thermal operations, is investigated.
We prove that with the help of a correlated catalyst, the state convertibility for any coherent state is fully characterized by the free energy defined with the quantum relative entropy.
We can extend this result to general resource theories in the form that imposing the covariant condition to a general resource theory does not change the state convertibility, as long as the initial state is coherent and distillable and this resource theory admits the phase estimation and the phase shift.
This means that adding a constraint from the law of energy conservation is irrelevant in the correlated-catalytic framework.
\end{abstract}

\maketitle

The central subject in quantum thermodynamics is the controllability of small quantum systems in a thermal environment.
The fundamental problems are what constraints can be seen in a small thermal environment and how robust the second law of thermodynamics is~\cite{Gour-review, Los-review, Sag-review}.
One standard implementation of this setup is an operational class of thermal operations, where we employ an energy-conserving unitary operation by consuming an auxiliary system at the Gibbs state~\cite{Jan00, HO13, Bra13}.
However, since thermal operations are formulated in a bottom-up fashion and are not easy to handle directly~\cite{NG, Los15, Cwi15, Gou18, FBB19}, it is convenient to focus on some key properties of thermal operations and examine them separately.

One key characterization of thermal operations is the Gibbs-preserving property that an initial Gibbs state results in the same Gibbs state~\cite{Bra15, LMP15, WNW19, SS21}.
Unlike conventional macroscopic thermodynamics, various novel constraints other than the conventional second law emerge in small systems~\cite{Bra15, AN, Kli, Tur, MOAbook, WNW19, FR18}.
However, interestingly, these novel constraints easily disappear under a small modification, and in various setups, the conventional second law of thermodynamics with a single free energy can be recovered in small quantum regimes~\cite{SSP13, SSP14, LMP15, Kor16, SS21, Mul18, Fai19, Sag21, Gou22}.
One of the standard setups for this recovery is that with a correlated catalyst~\cite{Dat22, BWN23, Shi25}:
In this framework, there is an auxiliary system $C$ whose reduced state does not change through the process while it helps the state conversion such that a state conversion $\rho\to \rho'$ is implemented by $\rho\otimes c\to \tau$ with $\Tr_S[\tau]=c$ and $\Tr_C[\tau]=\rho'$, where $c$ is the state of the catalyst.
In the resource theory with Gibbs-preserving operations, the state convertibility with a correlated catalyst is characterized by a single second law in both classical~\cite{Mul18} and quantum systems~\cite{SS21}.

Thermal operations have, however, another key characterization; the covariant condition, which reflects the restriction from the law of energy conservation.
The covariant condition is a genuinely quantum property imposing a constraint on the transformation of coherence among energy eigenstates:
We cannot convert, for example, an energy eigenstate into a superposition of two energy eigenstates without any additional help, which serves as a severe restriction on possible operations.
The Gibbs-preserving map and thermal operation coincide in the classical regime~\cite{HO13, Shi20}, while these two have a gap in the quantum regime due to the restriction on coherence~\cite{FOR15}.

The restriction from the covariant condition has been studied in the resource theory of asymmetry, or unspeakable coherence~\cite{Jan06, GS08, Mar-thesis, MS14, MS16}.
Previous studies on the resource theory of asymmetry with a correlated catalyst have revealed two contrastive faces.
If the initial state has no coherence, then the final state still has no coherence~\cite{LM19, MS19}.
On the other hand, if the initial state has maybe small but finite coherence, then a covariant operation can convert this state into any state including a maximally coherent state~\cite{ST23, KGS23}.
In other words, in covariant operations, whether coherence is zero or nonzero matters while the amount of coherence does not matter.

The Gibbs-preserving condition and the covariant condition are considered to be two main restrictions on thermal operations~\cite{Los-review}.
This motivates us to introduce a class of operations which are both covariant and Gibbs-preserving, also called the {\it enhanced thermal operation}~\cite{Cwi15, Gou18, DDH21, WT24} (see \fref{Benn}).
Although it is theoretically more tractable than thermal operation, the treatment of covariant Gibbs-preserving operations is still complicated since these two conditions restrict operations in a completely different manner.
In fact, the state convertibility with enhanced thermal operations in the single-shot regime obtained in Ref.~\cite{Gou18} is extremely complicated.
A clear understanding of the combination of covariance and Gibbs-preserving properties is still elusive.

In this Letter, we address this problem and establish the necessary and sufficient condition of state conversions for coherent states.
We prove that the state convertibility of the covariant Gibbs-preserving operation (enhanced thermal operation) with a correlated catalyst is fully characterized by a single free energy defined by the relative entropy as long as the initial state has nonzero coherence.
In other words, in the correlated-catalytic framework, the mere Gibbs-preserving operations and the covariant Gibbs-preserving operations have the same state conversion power, and the only exception is the incoherent case.
This serves as strong support for the conjecture on thermal operations raised in Ref.~\cite{KGS23}.
We emphasize that although the obtained condition is the composition of the conditions for covariant operations and Gibbs-preserving operations, this result including its proof is not trivial at all.
In fact, in the single-shot regime, the condition of state conversions for the covariant Gibbs-preserving operations~\cite{Gou18} is not the composition of the conditions for covariant operations~\cite{Mar-thesis} and Gibbs-preserving operations~\cite{SS21}.

The obtained result can be extended to a wide class of other resource theories.
We demonstrate that in the correlated-catalytic framework, the addition of the covariant condition does not change the state convertibility as long as the initial state is coherent and distillable and this resource theory admits the phase estimation and the phase shift.
The above result applies to, e.g., the resource theory of entanglement.
This result justifies our ignorance of the constraint from the law of energy conservation in these resource theories.

\figin{5cm}{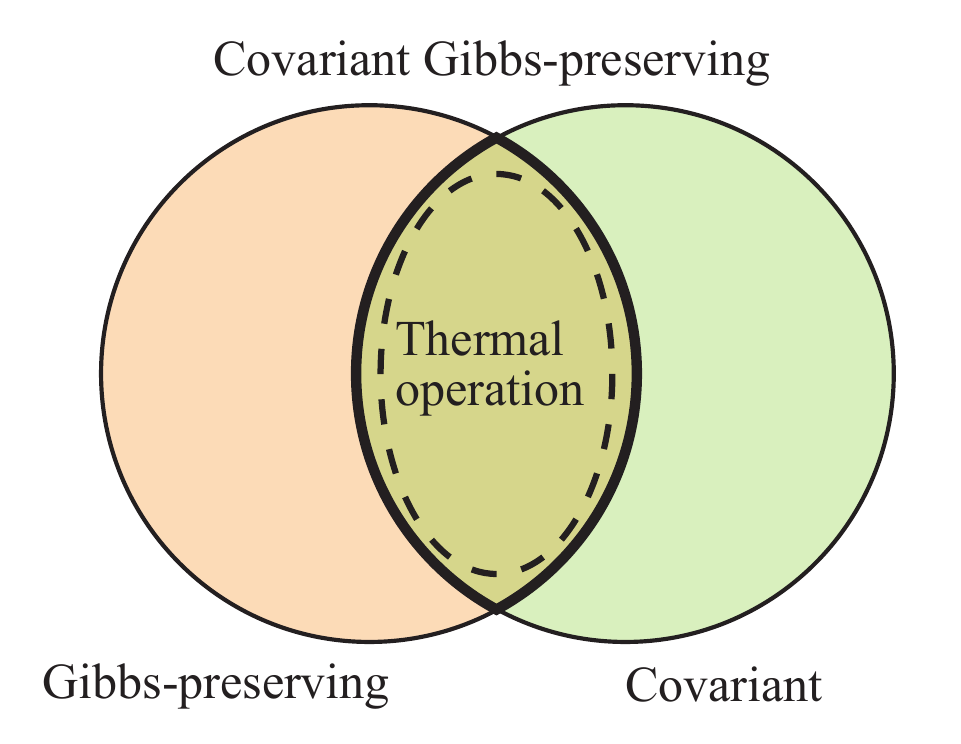}{
The relation of covariant Gibbs-preserving operations and thermal operations.
The covariant Gibbs-preserving operations (surrounded by a solid line) are the intersection of Gibbs-preserving operations and covariant operations, and thermal operations (surrounded by a dashed line) are its subset.
Whether it is a strict subset in the framework with a correlated-catalyst is an open problem.
}{Benn}

\para{Setup and main result}
Consider a system whose energy level spacings are integer multiples of some value $\Di$ (i.e., all level spacings are relatively rational).
This assumption is employed in various previous studies in the resource theory of (unspeakable) coherence~\cite{GS08, Abe14, Mar20, Mar22}, and thus we follow this convention.
Any actual system in experiments have ambiguity in the evaluation of the value of eigenenergy, and thanks to the denseness of rational numbers, within this ambiguity there exists a system with relatively rational level spacings (For more discussion related to this point, see the Supplemental Material~\cite{SM}).
Under this assumption, any state $\rho$ has period $\tau=2\pi /\Di$, i.e., $e^{-iH\tau}\rho e^{iH\tau}=\rho$.
Here, we normalize the Planck constant to unity.

A map $\calE: S\to S$ is called a covariant Gibbs-preserving operation (CGPO) if (i) $\calE(\rhoG)=\rhoG$ and (ii) $\calE(e^{-iHt}\rho e^{iHt})=e^{-iHt}\calE(\rho) e^{iHt}$ for any $t$ are satisfied.
Here, $\rhoG:=e^{-\beta H}/Z$ is a Gibbs state with a given inverse temperature $\beta$
The CGPO is also called enhanced thermal operation in Ref.~\cite{Cwi15}.

We investigate the correlated-catalytic conversion with CGPO in this Letter.
We say that $\rho$ is convertible to $\rho'$ through CGPO with a correlated-catalyst if for any $\ep>0$ there exist an auxiliary system $C$ called a catalyst, its state $c$, and a CGPO $\calE: S\otimes C\to S\otimes C$ such that $\tau=\calE(\rho\otimes c)$ with $\Tr_S[\tau]=c$ and $\abs{\Tr_C[\tau]-\rho'}_1<\ep$.
If the last condition is replaced by $\Tr_C[\tau]=\rho'$, we say this conversion {\it exact}.

Now we state our first main result, which serves as an important milestone toward a fully quantum thermodynamics with quantum coherence.
Here we define the free energy $F(\rho):=S(\rho||\rhoG)$ with the relative entropy $S$~\cite{Haybook}.
\bthm{\lb{t:main}
Consider two states $\rho$ and $\rho'$ in $S$.
We assume that the shortest period of $\rho$ is $2\pi/\Di$ (i.e., all modes are coherent).
Then, $\rho$ is convertible to $\rho'$ through CGPO with a correlated-catalyst if and only if $F(\rho)\geq F(\rho')$.
In addition, if $F(\rho)>F(\rho')$ and $\rho'$ is full-rank, then this conversion is exact.
}
This theorem clearly shows the recovery of the second law of thermodynamics in CGPO as long as the initial state has coherence.
This is the first result establishing the universality of the second law of thermodynamics under the covariant operation without any consumption of coherence in external systems.
As clearly demonstrated by this theorem, the presence of quantum coherence does not disturb thermodynamic state conversions, if small but finite coherence exists at the initial stage.
This shows a clear contrast to previous suggestions that the quantum coherence serves as a severe constraint in quantum thermodynamics~\cite{Cwi15, Gou18, NG, Los15}.

To prove this theorem, a special type of asymptotic conversion, called the {\it marginal-asymptotic conversion}~\cite{Fer23, GKS23}, plays a pivotal role.
We say that $\rho$ is convertible to $\rho'$ in a marginal asymptotic conversion through CGPO with conversion rate $r$ with arbitrarily small error if for any $\ep>0$ there exists a sufficiently large integer $N$ and a CGPO $\Lambda: S^{\otimes N}\to S'^{\otimes \lfloor rN\rfloor}$ such that $\Xi=\Lambda(\rho^{\otimes N})$ with $\abs{\Tr_{\bcs i}[\Xi]-\rho'}_1<\ep$ for all $i=1,2\ldots , \lfloor rN\rfloor$.
The difference between the conventional asymptotic conversion and the marginal-asymptotic conversion lies in the fact that we measure the error by the final state of the entire copies in conventional asymptotic conversions, while by reduced states of each copy in marginal-asymptotic conversions.
This apparently small difference in definitions is in fact so important that the following lemma, which is a key step of the proof of \tref{main}, is valid only for marginal-asymptotic conversions.
\blm{\lb{t:lemma}
Assume the same assumptions as \tref{main} and $F(\rho)\geq F(\rho')$.
Then, for any $\delta >0$ there exists a marginal-asymptotic conversion from $\rho$ to $\rho'$ through CGPO with conversion rate $1-\delta$ with arbitrarily small error.
}
It is known that an approximate asymptotic conversion can be transformed into a correlated-catalytic conversion~\cite{SS21, TS22, KGS23} (see also \cite{Wil21, LS21, KDS21, LJ21, Cha21, Wil22, RT22, LN23} for its applications).
Using this connection, \lref{lemma} directly implies the existence of a correlated-catalytic conversion from $\rho$ to $\rho'$ with arbitrarily small error; \tref{main}.

\para{Extension to general resource theories}
As will be seen soon later, the proof of \tref{main} does not utilize detailed properties of the Gibbs-preserving property.
What we use is the fact that (a) distillation is possible by a GPO, and (b) the phase estimation and the phase shift can be performed by a GPO.
Using these facts, the problem is reduced to whether a marginal asymptotic conversion from $\rho$ to $\rho'$ with rate 1 exists in GPO.

Thus, if the above two conditions are satisfied by a resource theory $\bbO$, our proof technique also works for this resource theory, and state convertibility in this resource theory with adding the covariant condition $\bbO\cap {\rm Cov}$ is fully determined whether a marginal-asymptotic conversion from $\rho$ to $\rho'$ with rate 1 exists in $\bbO$.
In this situation, whether the desired marginal-asymptotic conversion exists is the central problem.

Fortunately, a recent result by Ganardi {\it et al}.~\cite{GKS23} reveals the equivalence of a marginal-asymptotic conversion with rate 1 and a correlated-catalytic conversion for distillable states.
Applying this result, we conclude that if the initial state is distillable and coherent, and this resource theory admits the phase estimation and the phase shift, then the state convertibility does not change by further imposing the covariant condition.
This is the second main result of this Letter.
\bthm{\lb{t:gen}
Consider a resource theory whose free operations are $\bbO$.
Suppose that $\rho$ is convertible to $\rho'$ with a correlated catalyst, $\rho$ is distillable, and $\rho$ has period $2\pi/\Delta$.
In addition, the phase estimation and the phase shift can be performed in $\bbO$.
Then, in a resource theory with free operations, $\bbO\cap {\rm Cov}$, $\rho$ is convertible to $\rho'$ with a correlated catalyst.
}
Surprisingly, the covariant condition, which forces us to pay attention to quantum coherence, puts no restriction on state convertibility in a wide class of resource theories.
In the investigation of resource theories, we usually ignore the constraint from energy conservation, though the law of energy conservation is universal, and thus any resource theory should take this restriction into account.
This striking theorem justifies this ignorance by showing that a resource theory with and without the covariant condition has essentially the same state conversion power in the framework with a correlated-catalyst.
Let us take as an example the resource theory of entanglement, where the set of free operations $\bbO$ is LOCC (local operations and classical communications)~\cite{Haybook}.
In this case, the true set of accessible operations in our world is LOCC $\cap$ Cov (local operations and classical communications under energy conservation).
However, by recalling that a state tomography is possible by LOCC, and thus the phase estimation is tractable in the framework with LOCC, it suffices to treat simply the class of LOCC, since LOCC and LOCC $\cap$ Cov provide the same state convertibility for all distillable states with finite coherence.


\para{Conversion protocol for CGPO}
Below, we construct a CGPO protocol realizing the desired state conversion claimed in \lref{lemma}.
To construct our protocol, we employ the following two established facts.
One is on a GPO, stating that state convertibility by a GPO in the conventional asymptotic conversion is fully characterized by the free energy.
\bpro{[Gibbs-preserving map (used in \cite{Fai19, SS21})]\lb{t:GPO}
Consider two states $\rho$ and $\rho'$ in $S$ with $F(\rho)\geq F(\rho')$.
Then, for any $\delta_1>0$ there exists a large $M$ and a GPO $\Lambda: S^{\otimes M}\to S^{\otimes M}$ such that $\abs{\Lambda(\rho^{\otimes M})-\rho'^{\otimes M}}_1<\delta_1$.
}

The other is on a covariant operation, stating that we can efficiently estimate the phase of a state by a covariant operation.
Here, the phase (time) estimation is a task to evaluate $\tau$ from many copies of $e^{-iH\tau}\rho e^{iH\tau}$.
\bpro{[Covariant time estimation (used in \cite{Mar20})]\lb{t:cov}
There exists a time estimation protocol $S^{\otimes m}\to \bbR$ with probability distribution $P(\test|\kappa)$ for $\kappa$ on $S^{\otimes m}$ such that (i) covariant: $P(\test|e^{-iH^{\otimes m}\tau}\kappa e^{iH^{\otimes m}\tau})=P(\test+\tau|\kappa)$ for any $\kappa$ and $\tau$, and (ii) the variance of $P(\test|\rho^{\otimes m})$ decays as $O(1/m)$ for any $\rho$ in $S$ whose shortest period is $2\pi/\Di$.
}
Without loss of generality, we suppose that the average of $\test$ for $\rho$ is zero (i.e., we set the origin of the phase as supposition).

\figin{8.5cm}{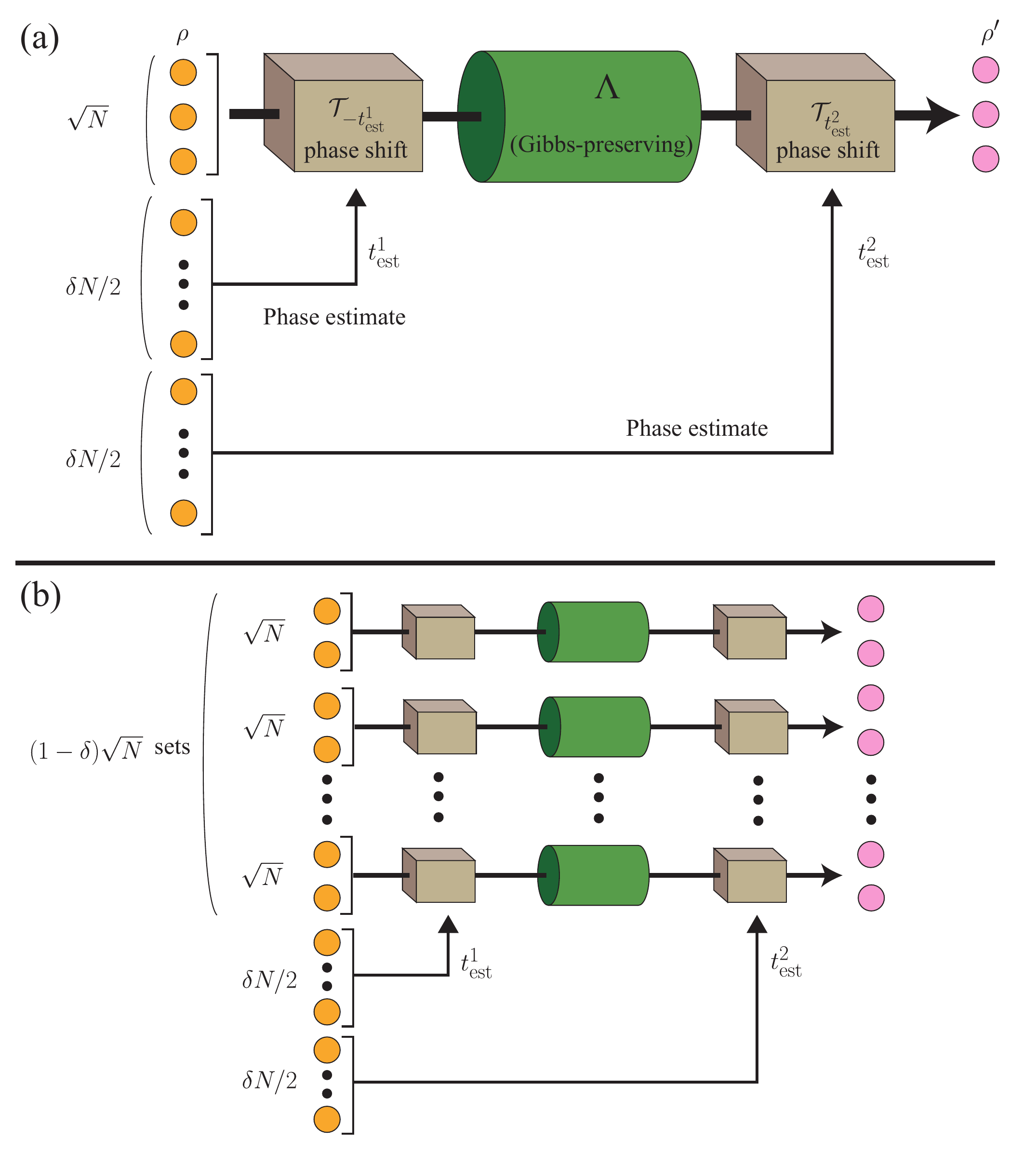}{
(a): A protocol for a single set $A_i$ converting $\rho^{\otimes \sqrt{N}}$ to $\rho'^{\otimes \sqrt{N}}$.
We estimate the times $\test^1$ and $\test^2$ by using $\delta N/2$ copies of $\rho$, respectively, and apply the phase shift before and after the Gibbs-preserving operation.
(b): The whole protocol of the marginal-asymptotic conversion from $\rho$ to $\rho'$ with transformation rate $1-\delta$.
Here, all the phase shift protocols employ the same estimaters $\test^1$ and $\test^2$.
}{protocol}

Now we construct a protocol converting $N$ copies of $\rho$ into $(1-\delta)N$ copies of $\rho'$ for sufficiently large $N$.
We decompose $N$ copies of $\rho$ into $\nu:=(1-\delta)\sqrt{N}$ sets of $\sqrt{N}$ copies and two $\delta N/2$ copies, which we call $A$ part, $B_1$ part, and $B_2$ part respectively.
The sets in $A$ part are labeled as $A_1,\ldots , A_\nu$.
In addition, for state $R$ on $S^{\otimes n}$, we introduce a time propagator by $t$ defined as
\eq{
\calT_t(R):=e^{-iH^{\otimes n}t} R e^{iH^{\otimes n}t}.
}
In the following, we choose $n$ properly depending on state $R$.

Now we construct the marginal-asymptotic conversion protocol which consists of two steps:
\be{
\item We apply the time estimation protocol of Prop.~\ref{t:cov} on $B_1$ part and $B_2$ part (two $\delta N/2$ copies of $\rho$) and obtain $\test^1$ and $\test^2$.
The variance of $\test$ decays as $O(1/\delta N)$.
\item We apply the map
\eq{
\calM( \rho^{\otimes \nu\sqrt{N}}):=\calT_{\test^2}\circ \Lmd^{\otimes \nu} \circ \calT_{-\test^1} (\rho^{\otimes \nu \sqrt{N}})
}
on each set $A_i$.
Here, $\Lambda$ is the Gibbs preserving map converting $\rho^{\sqrt{N}}$ to $\rho'^{\sqrt{N}}$, whose existence is shown in Prop.~\ref{t:GPO}.
}

It is not hard to demonstrate that this protocol is indeed covariant, Gibbs-preserving, and realizes the desired conversion (see Supplemental Material~\cite{SM}).
The covariant condition is demonstrated as follows:
Consider an input state in the form of $e^{-iHt}\rho e^{iHt}$.
The first phase estimator $\test^1$ cancels the phase shift in the initial state, which guarantees that the input state for $\Lambda$ is always the same regardless of the initial phase $t$.
Then, the second phase estimator $\test^2$ recovers the phase shift, which fulfills the covariant condition.
Note that we need to perform time estimation twice since the reuse of estimator $\test$ accompanies unwanted correlation between two phase-shift operations and the covariant condition becomes unsatisfied.

Some complicated construction, e.g., decomposition into sets with $\sqrt{N}$ copies, is necessary to realize the desired transition with small errors.
\pref{cov} implies that $\calT_{\test}$ on $a$ copies of $\rho$ with $\test$ estimated from $b$ copies of $\rho$ has phase variance of order $\frac ab$.
Since we need to vanish the phase variance, the number of copies in each set ($\sqrt{N}$ in our case) should be sublinear in the number of copies used for phase estimation ($\delta N/2$ in our case).
In addition, to realize the conversion rate to be close to 1, the number of copies should satisfy the following order relation: (copies in a single set)$\ll$(copies for phase estimation)$\ll$(total converted copies), which is chosen so that negligibly few copies are consumed for phase estimations compared to the converted copies to $\rho'$.




\para{Comments and perspectives}
We established the necessary and sufficient condition of state conversions with covariant Gibbs-preserving operations (enhanced thermal operations) in the correlated-catalytic framework.
Our result confirms the recovery of the second law of thermodynamics in a fully quantum regime under the law of energy conservation, and the only exception is the incoherent case, whose measure in the state space is zero.
The constraint from the energy conservation only separates the coherent case and the incoherent case, and no further restriction is imposed on coherent states.
This structure is not special to quantum thermodynamics but widely seen in general resource theories.
Namely, we demonstrated that the addition of the covariant condition does not change the state convertibility, as long as the state is coherent and distillable and this resource theory admits the phase estimation and the phase shift.
This fact supports why various resource theories work well without taking into account the universal and evidently present constraint from the law of energy conservation.

We note that the present result applies only to systems whose energy level spacings are relatively rational, which is necessary for our present phase estimation processes.
This limitation has the same route as the sublinear asymptotic conversion protocol under the covariant condition proposed by Marvian~\cite{Mar20}, which can be diverted into a marginal asymptotic conversion protocol~\cite{Shi25}.
To overcome this limitation, it appears fruitful to employ a recent result on the triviality of the resource theory of (unspeakable) coherence with a correlated catalyst~\cite{ST23, KGS23}, which applies to general systems with relatively irrational energy level spacings.
We strongly expect that the same result holds for general systems, though combining the above result and the results presented in this Letter seems not straightforward.

An interesting question is the connection to thermal operations, which is the most common and physically supported class of free operations in quantum thermodynamics.
The Gibbs-preserving condition and the covariant condition are the two central conditions of thermal operations, and whether these two are the only relevant conditions is an open problem~\cite{Los-review}.
If the answer to the above question is positive, then the covariant Gibbs-preserving operations fully capture the structure of thermal operations, and the conjecture raised in Ref.~\cite{KGS23} that the quantum thermodynamics with thermal operations is characterized by a single free energy for coherent states is resolved in the affirmative.
At present, our result provides a consistent view to this conjecture.
We note that the covariant Gibbs-preserving operations and thermal operations are known to have a gap in the single-shot regime~\cite{Cwi15, DDH21}.
However, since many inequivalent resource theories collapse in the correlated-catalytic framework~\cite{SN23}, it is still plausible to expect that the gap between the covariant Gibbs-preserving operations and thermal operations is unstable and these two have the same conversion law in the correlated-catalytic framework.
Whether these two are indeed equivalent or not is left as an open problem to be solved.

{\it Acknowledgments.} ---
The author thanks Ryuji Takagi for the fruitful discussion.
This work was supported by JST ERATO Grant Number JPMJER2302, Japan.


\let\oldaddcontentsline\addcontentsline
\renewcommand{\addcontentsline}[3]{}

\let\addcontentsline\oldaddcontentsline

\clearpage

\pagestyle{plain}

\setcounter{page}{1}
\setcounter{figure}{0}
\renewcommand{\thefigure}{S.\arabic{figure}}

\makeatletter
\long\def\@makecaption#1#2{{
\advance\leftskip1cm
\advance\rightskip1cm
\vskip\abovecaptionskip
\sbox\@tempboxa{#1: #2}%
\ifdim \wd\@tempboxa >\hsize
 #1: #2\par
\else
\global \@minipagefalse
\hb@xt@\hsize{\hfil\box\@tempboxa\hfil}%
\fi
\vskip\belowcaptionskip}}
\makeatother
\newcommand{\vo}{\upsilon}
\newcommand{\midskip}{\vspace{3pt}}

\setcounter{thm}{0}
\renewcommand{\thethm}{S.\arabic{thm}}
\renewcommand{\thedfn}{S.\arabic{dfn}}
\renewcommand{\thelm}{S.\arabic{lm}}
\renewcommand{\thepro}{S.\arabic{pro}}


\onecolumngrid


\begin{center}
{\large \bf Supplemental Material for  \protect \\ 
  ``Quantum thermodynamics with coherence: Covariant Gibbs-preserving operation is characterized by the free energy'' }\\
\vspace*{0.3cm}
Naoto Shiraishi$^{1}$ \\
\vspace*{0.1cm}

{$^{1}${\it Department of Basic Science, The University of Tokyo} } 
\end{center}

\setcounter{equation}{0}
\renewcommand{\theequation}{S.\arabic{equation}}

In this Supplemental Material, we first present a brief review on quantum thermodynamics and relevant results in resource theories which we use in the proof of our main results.
We then provide proofs of \tref{main}, \lref{lemma}, and \tref{gen}.
All the reference numbers are common to the main text.

\tableofcontents

\section{Background}

\subsection{Quantum thermodynamics}\lb{s:QT-review}

In quantum thermodynamics, we investigate state convertibility between two given quantum states $\rho$ and $\rho'$ by {\it thermodynamic processes}.
Since we know the universality and robustness of the second law of thermdynamics, in the investigation in quantum thermodynamics we expect the emergence of the second law of thermodynamics in the small quantum regime with a proper physical setup.

We start with the characterization of thermodynamic processes.
There are at least two fashions to define what is thermodynamic processes:
One is thermal operations (TO), which is defined in a bottom-up manner.
and the other is Gibbs-preserving operation (GPO), which is defined in an axiomatic manner.
In the following, we denote the Hamiltonian of system $S$ by $H$, and the inverse temperature of environment $\beta$ is provided.

\bdf{[Thermal operation (TO)]
A CPTP map $\calE: S\to S$ is called a thermal operation (TO) if there exists a proper auxiliary system $B$ with Hamiltonian $H_{\rm B}$ and an energy-conserving unitary $U$ acting on the composite system $SB$ (i.e., $U$ commutes with $H+H_{\rm B}$) such that 
\eq{
\calE(\rho)=\Tr_B[U(\rho\otimes \rhoG^{\rm B})U^\dagger].
}
}

\bdf{[Gibbs-preserving operation (GPO)]
A CPTP map $\calE: S\to S$ is called a Gibbs-preserving operation (GPO) if $\calE$ maps the Gibbs state $\rhoG=e^{-\beta H}/Z$ to the same Gibbs state:
\eq{
\calE(\rhoG)=\rhoG.
}
}

If a map converts a state in system $S$ to another system $S'$ with Hamiltonian $H'$, GPO is defined as a map satisfying $\calE(\rhoG)=\rhopG$ with $\rhopG=e^{-\beta H'}/Z'$.

The TO is defined in an operational manner.
In contrast, the GPO is defined in a top-down, or axiomatic manner.
A top-down axiomatic definition is easier to analyze from a mathematical perspective, and thus many fruitful relations are first found in resource theories with axiomatic characterizations.
On the other hand, a bottom-up operational definition provides a clear physical picture how to implement this map.
In many resource theories, researchers frequently analyze a theory with a top-down characterization first and then move to theories with more bottom-up characterizations.
This is also true for quantum thermodynamics.

It is easy to show that a TO is also a GPO, that is, the class of GPO is larger than TO.
In addition, in the classical case (i.e., we only treat diagonalized states with respect to energy eigenstates), its opposite is also true, and these two classes are equivalent~\cite{HO13, Shi20}.
On the other hand, in the quantum case, the set of TO is known to be a strict subset of the set of GPO, that is, there exists a GPO which cannot be implemented by any TO~\cite{FOR15}.
An example is an operation creating a coherence among different energy eigenstates as $\ket{E_0}\to \alpha \ket{E_0}+\gamma \ket{E_1}$ ($\abs{\alpha}^2+\abs{\gamma}^2=1$).
This conversion can be Gibbs-preserving with a proper $\alpha$ and $\gamma$, while TO (an energy-conserving unitary with an auxiliary incoherent system) never realizes this state conversion.

From this observation, we notice that the barrier of quantum coherence coming from the law of energy conservation is another key characterization of TO.
This subject has already been treated in the resource theory of (unspeakable) coherence (or the resource theory of asymmetry).
In this context, an energy-conserving unitary with an incoherent auxiliary system is called a {\it covariant operation}.

\bdf{[Covariant operation]
A CPTP map $\calE: S\to S$ is called a covariant operation if 
\eq{
\calE(e^{-iHt}\rho e^{iHt})=e^{-iHt}\calE(\rho) e^{iHt}
}
is satisfied for any $t$.
}
\bdf{[Incoherent state]
A state $\rho$ is incoherent if 
\eq{
e^{-iHt}\rho e^{iHt}=\rho
}
is satisfied for any $t$.
}

The following theorem confirms that both axiomatic and operational characterizations coincide.
The proof of this equivalence is presented in e.g., Ref.~\cite{Mar-thesis}.

\bpro{
For any covariant operation $\calE$, there exists a proper auxiliary system $A$ with Hamiltonian $H_{A}$, an incoherent state $\xi$ on $A$, and an energy-conserving unitary $U$ on the composite system $SA$ such that 
\eq{
\calE(\rho)=\Tr_A[U(\rho\otimes \xi)U^\dagger].
}
}

\bigskip

The Gibbs-preserving property and the covariant property are considered to be two key properties of TO~\cite{Los-review, Shi25}.
Thus, to address TO, it is reasonable to treat a GPO and a covariant operation separately in the first stage, and then combine these two results.
After them, we finally examine TO itself.
Along this story, this section reviews results in the first stage, and our main theorems investigate the second step, their combination.

\subsection{Catalytic frameworks and asymptotic frameworks}

In resource theories, we restrict a class of possible operations we can address and investigate state convertibility by these operations.
The set of accessible operations is called {\it free operations}, which we denote by $\bbO$.

When we use no extra help in conversion by a free operation and simply consider conversion $\rho\to \rho'$ by a free operation in $\bbO$, we call this problem setting {\it single-shot}.
Although the shingle-shot state convertibility is studied in various resource theories, the power of single-shot state conversion is usually weak, and therefore some extended setups enhancing the power of state conversion are frequently adopted.
On the theoretical side, the necessary and sufficient condition of state convertibility in the single-shot regime usually contains extremely many conditions, and most of them are not robust under a certain small perturbations.
Extracting only robust conditions is physically meaningful, and through this we usually reach a mathematically simple structure.

\subsubsection{Catalytic frameworks}\lb{s:catalytic}

One standard framework in resource theories is the catalytic framework, where we employ an auxiliary system called {\it catalyst}~\cite{AHJ18, Dat22, BWN23, Shi25}.
The catalyst does not change its own state through the conversion process, while it helps the state conversion in the system.
We here introduce two types of catalysts; an uncorrelated catalyst and a correlated catalyst.
The uncorrelated catalyst does not have correlation between the system and the catalyst at the final state.
In contrast, the correlated catalyst is allowed to have a small correlation between the system and the catalyst, and the reduced state of the catalyst returns to its original state.

\bdf{[uncorrelated catalyst]
We say that $\rho$ is convertible to $\rho'$ by a class of free operations $\bbO$ with an uncorrelated catalyst with a vanishing error if for any $\ep>0$ there exist a catalyst $C$, its state $c$, and a free operation $\Lambda\in \bbO: SC\to SC$ such that 
\eq{
\Lambda(\rho\otimes c)=\kappa\otimes c
}
with $\abs{\kappa-\rho'}_1<\ep$.
}

\bdf{[correlated catalyst]
We say that $\rho$ is convertible to $\rho'$ by a class of free operations $\bbO$ with a correlated catalyst with a vanishing error if for any $\ep>0$ and $\delta>0$ there exist a catalyst $C$, its state $c$, and a free operation $\Lambda\in \bbO: SC\to SC$ such that 
\eq{
\Lambda(\rho\otimes c)=\tau
}
with $\Tr_S[\tau]=c$, $\abs{\Tr_C[\tau]-\rho'}_1<\ep$, and $\abs{\tau-\Tr_C[\tau]\otimes c}_1<\delta$.
}

Here, $\abs{A}_1:=\Tr[\sqrt{A^\dagger A}]$ is the trace norm~\cite{Haybook}.

Note that the above definitions employ  {\it approximate conversion with a vanishing error}, meaning that the error $\ep$ is finite but can be set arbitrarily small by preparing a proper catalyst.
Aside from this, we can also define catalysts for {\it exact conversion}.
For comparison, we present these definitions below:

\bdf{[uncorrelated catalyst (exact case)]
We say that $\rho$ is convertible to $\rho'$ by a class of free operations $\bbO$  with an uncorrelated catalyst exactly if there exist a catalyst $C$, its state $c$, and a free operation $\Lambda\in \bbO: SC\to SC$ such that 
\eq{
\Lambda(\rho\otimes c)=\rho'\otimes c.
}
}

\bdf{[correlated catalyst (exact case)]\lb{def-cc-exact}
We say that $\rho$ is convertible to $\rho'$ by a class of free operations $\bbO$  with a correlated catalyst exactly if for any $\delta>0$ there exist a catalyst $C$, its state $c$, and a free operation $\Lambda\in \bbO: SC\to SC$ such that 
\eq{
\Lambda(\rho\otimes c)=\tau
}
with $\Tr_S[\tau]=c$, $\Tr_C[\tau]=\rho'$, and $\abs{\tau-\rho'\otimes c}_1<\delta$.
}

The notion of the catalyst is understood analogously to axiomatic characterization of conventional thermodynamics~\cite{LY99}.
Thermodynamics can be described by the convertibility of two equilibrium states $X$ and $X'$ {\it without leaving any change in the remaining of the universe}.
The last qualification means that we can borrow and return a state from elsewhere in the universe, which corresponds to the role of the catalyst.
In fact, if the borrowed auxiliary system goes back to the original state, we cannot distinguish whether this system is borrowed or not.
Note that the universe has extremely various states, and thus we expect that any desired auxiliary system can be found somewhere.
Thus, to put a fundamental limitation of state conversions, the presence of catalysts is usually taken into account regardless of the size and complexity of catalysts.
The small correlation in the case of a correlated catalyst is also understood analogously to the observation in thermodynamics that small correlation between the system and the external environment is regarded to be negligible.

\bigskip

We shall show how state convertibility changes among the frameworks, single-shot (without catalyst), with an uncorrelated catalyst, and with a correlated catalyst, by taking as an example the GPO in the classical regime, which is presented in \sref{GPO-classical} (\pref{Blackwell} (no catalyst), \pref{second-laws} (uncorrelated catalyst), and \pref{c-second-law} (correlated catalyst)).
Below we summarize these three results:
\bi{
\item \ul{(single-shot; no catalyst)}: An intuitive but not simple graphical condition is the necessary and sufficient condition of state conversion.
\item \ul{(uncorrelated catalyst)}: A family of infinite second-law-like inequalities  is the necessary and sufficient condition of state conversion.
\item \ul{(correlated catalyst)}: The second law of thermodynamics is the necessary and sufficient condition of state conversion.
}

As clearly seen from above, the conditions of state conversion become simpler in order of without catalyst, with an uncorrelated catalyst, and with a correlated catalyst.
In particular, we succeed in recovering the second law of thermodynamics in the framework with a correlated catalyst.

\bigskip

We finally introduce a useful technique which transforms results on approximate conversions with a vanishing error into those on exact conversions.
This technique was first shown in \cite{Wil22}.

The core of this technique is a genuine mathematical fact on convex sets.
Here, we presuppose that a proper distance between quantum states is defined.

\bpro{[Ref.~\cite{Wil22}]\lb{t:Wil}
Consider a sequence of convex sets of quantum states $\{S_m\}_{m=1}^\infty$ satisfying $S_m\subseteq S_{m+1}$ for any $m$.
Let $V$ be the closure of $\lim_{m\to \infty} S_m$.
If $\kappa$ is an interior state of $V$, then there exists an integer $m$ such that $\kappa\in S_m$.
}

Now we explain how to use this proposition to catalytic conversions.
Consider a resource theory with a set of free operations $\bbO$, and choose $V$ as a set of states convertible from $\rho$ with a (correlated or uncorrelated) catalyst with a vanishing error.

We set $S_m$ as a set of states convertible from $\rho$ exactly through a free operation in $\bbO$ with a catalyst whose Hilbert space has a dimension less than or equal to $m$.
If a classical mixture process is in $\bbO$, then $S_m$ is a convex set.
Hence, if $\rho'$ is an interior point of $V$, \pref{Wil} suggests $\rho'\in S_m$ for some $m$.
This means that $\rho'$ is obtained from $\rho$ with a catalyst with dimension $m$ exactly.

\bpro{[Transformation from approximate conversion with vanishing error to exact conversion~\cite{Wil22, TS22}]\lb{t:approximate-exact}
Consider a resource theory with a set of free operation $\bbO$ which includes a classical mixture process.
We denote by $V_\rho$ the set of states achievable from $\rho$ by $\bbO$ with a (correlated or uncorrelated) catalyst with a vanishing error.
Then, if $\rho'$ is an interior point of $V_\rho$, then $\rho$ is convertible to $\rho'$ by $\bbO$  with a (correlated or uncorrelated) catalyst exactly.
}

\subsubsection{Asymptotic conversions}

Another standard framework in resource theories is the asymptotic framework, where we consider conversions of many copies of input state $\rho$ to many copies of output state $\rho'$ (i.e., $\rho^{\otimes n}\to \rho^{\otimes m}$).
In the asymptotic framework, the conversion rate $r=\lim_{n\to\infty}\frac mn$ is a key quantifier.
A large rate means that we can obtain many $\rho'$'s from a small number of $\rho$'s, implying that $\rho$ is more resourceful than $\rho'$.

Similarly to the catalytic framework, we usually allow a vanishing error in the final state.
There are at least two manners to measure the amount of the error, which lead to asymptotic conversions and marginal asymptotic conversions, respectively.

\bdf{[Asymptotic conversion]
We say that $\rho$ is convertible to $\rho'$ by a marginal asymptotic conversion through an operation class $\bbO$ with conversion rate $r$ with an arbitrarily small error if for any $\ep>0$ there exists a sufficiently large integer $N$ and a map $\Lambda\in\bbO: S^{\otimes N}\to S^{\otimes \lfloor rN\rfloor}$ such that $\Xi=\Lambda(\rho^{\otimes N})$ with 
\eq{
\abs{\Xi -\rho'^{\otimes \lfloor rN\rfloor}}_1<\ep.
}
}

\bdf{[Marginal-asymptotic conversion]
We say that $\rho$ is convertible to $\rho'$ by a marginal asymptotic conversion through an operation class $\bbO$ with conversion rate $r$ with an arbitrarily small error if for any $\ep>0$ there exists a sufficiently large integer $N$ and a map $\Lambda\in\bbO: S^{\otimes N}\to S^{\otimes \lfloor rN\rfloor}$ such that $\Xi=\Lambda(\rho^{\otimes N})$ with 
\eq{
\abs{\Tr_{\bcs i}[\Xi]-\rho'}_1<\ep
}
for all $i=1,2\ldots , \lfloor rN\rfloor$.
}

In the conventional asymptotic conversions, the error in the final state is measured by using all copies.
In contrast, in the marginal-asymptotic conversions, the error in the final state is measured by using each single copy.
Owing to the monotonicity of the trace norm, all asymptotic conversions are also marginal asymptotic conversions, while the opposite direction is not true in general.

The marginal-asymptotic conversion is introduced in Ref.~\cite{Fer23} and investigated further in Refs.~\cite{GKS23, ST23, KGS23}.
The definition of marginal-asymptotic conversions appears slightly artificial, while it is useful to proving the existence of other state conversions as shown in the next subsection.

\subsubsection{Connection between catalytic frameworks and asymptotic frameworks}\lb{s:catalytic-asymptotic}

The introduced two frameworks, the catalytic framework and the asymptotic framework, are not independent but have a deep connection with each other.

We first demonstrate that the existence of a marginal-asymptotic conversion with conversion rate 1 implies the existence of a correlated-catalytic conversion.

\bpro{[Refs.~\cite{SS21, TS22, KGS23}]\lb{t:asymptotic-catalytic}
Consider a resource theory whose free operations $\bbO$ include the relabeling of a classical register and the conditioning of free operations by a classical register.
Suppose that $\rho$ is convertible to $\rho'$ by a marginal-asymptotic conversion in $\bbO$ with conversion rate 1 (i.e., for any $\ep>0$ there exists a free operation $\Lambda: S^{\otimes n}\to S^{\otimes n}$ in $\bbO$ which maps $\rho^{\otimes n}$ to $\tau$ with $\abs{\rho'-\Tr_{\bcs i}[\tau]}<\ep$ for any $i=1,\ldots , n$).
Then, there exists a correlated-catalytic conversion in $\bbO$ which converts $\rho$ to $\rho'$ with error $\ep>0$.
}

\figin{14cm}{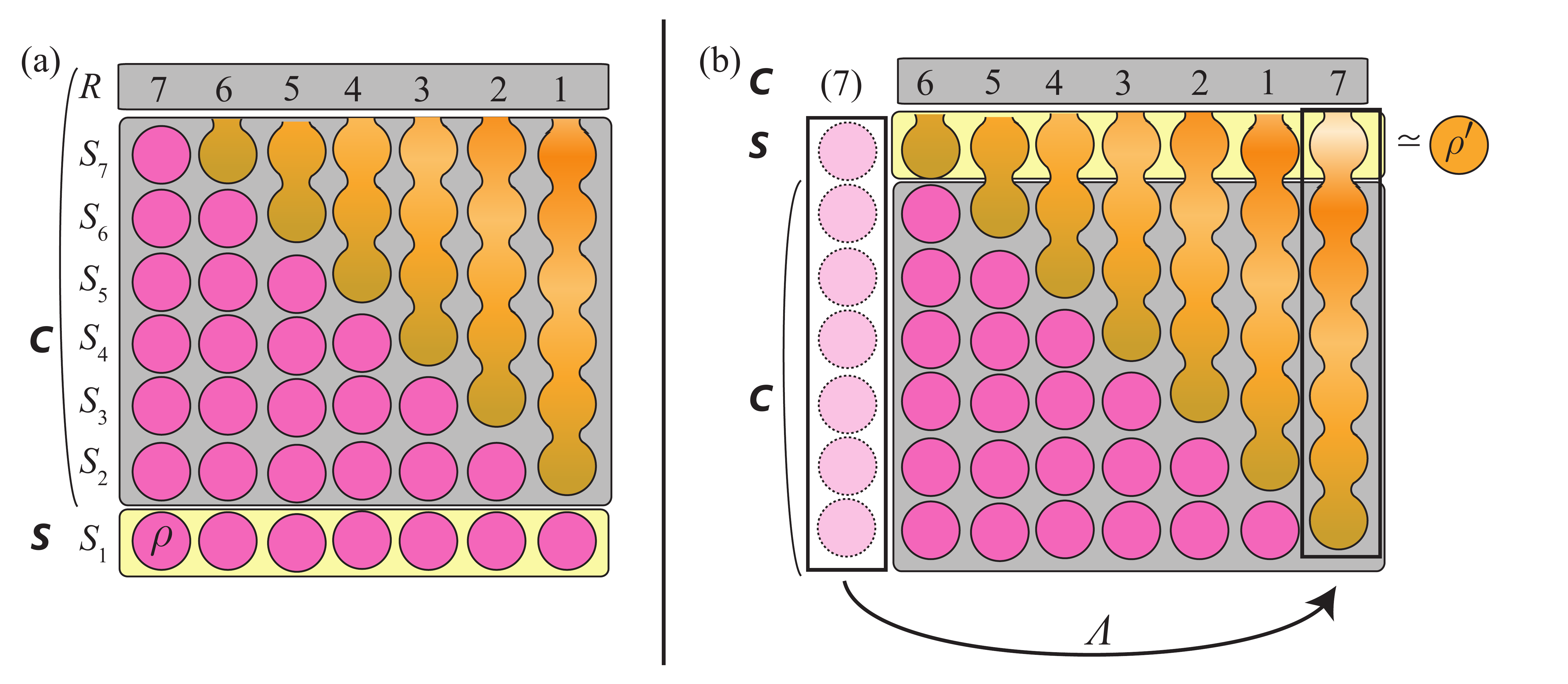}{
\justifying
An example of the transformation from a (marginal-)asymptotic conversion of 7 copies to a correlated-catalytic conversion.
(a): The structure of the catalyst given in \eref{catalyst-construction}.
Here, $R$ is a label system, and the horizontal direction means their classical mixture with equal weights.
In this case, we mix 7 states on $S^{\otimes 7}\otimes R$. 
(b):~The state after the application of the asymptotic conversion $\Lambda$.
By relabeling the system $S$ and the catalyst $C$, we find that the state of the catalyst recovers to its original state, and the system becomes close to the desired state $\rho'$.
}{asym-cat}

We here explain how to construct a correlated-catalytic conversion from a given marginal-asymptotic conversion with conversion rate 1, which we denote by $\Lambda$.
Below we construct the desired catalyst explicitly.
Let $\tau_i$ ($1\leq i\leq n$) be a reduced state of $\tau=\Lambda(\rho^{\otimes n})$ onto $S_1\otimes \cdots \otimes S_i$.
Using an $n$-dimensional label system $R$ whose Hamiltonian is trivial (i.e., all energies are degenerated), we construct the catalyst $c$ on $S^{\otimes n-1}\otimes R$ as follows (see \fref{asym-cat}.(a)):
\eqa{
c:=\frac1n \sum_{k=1}^n \rho^{\otimes k-1}\otimes \tau_{n-k}\otimes \ket{k}\bra{k}.
}{catalyst-construction}
Here, $\ket{k}$ is a state of the label system $R$.
The initial state of $SC$ reads 
\eq{
\rho\otimes c=\frac1n \sum_{k=1}^n \rho^{\otimes k}\otimes \tau_{n-k}\otimes \ket{k}\bra{k}.
}
By applying $\Lambda$ if the label system is $\ket{n}$ and relabeling the label system as $i\to i-1$ and $1\to n$, the state becomes (see \fref{asym-cat}.(b))
\eq{
\tau'=\frac1n \sum_{k=1}^n \rho^{\otimes k-1}\otimes \tau_{n+1-k}\otimes \ket{k}\bra{k}.
}
By regarding $n$ copies of states such that the last system is $S$ and the remaining is $C$, the catalyst returns to the original state $c$ and the state of the system becomes $\frac1n \sum_{k=1}^n \Tr_{\bcs k}[\tau]$, which is close to $\rho'$.

\bigskip

We next demonstrate that  the existence of a correlated-catalytic conversion implies the existence of a marginal-asymptotic conversion with conversion rate 1 if the initial state is distillable.
Here, a state is distillable if sufficiently many copies of this state can be converted to any (single) state by a free operation with a vanishing error.

\bdf{[Distillable state]
Consider a resource theory with a set of free operations $\bbO$.
In this resource theory, a state $\rho$ is distillable if for any $\rho'$ there exists a proper integer $\mu \in \bbN_+$ and a free operation $S^{\otimes \mu}\to S$ converting $\rho^{\otimes \mu}$ to $\rho'$ with a vanishing error.
}

\bpro{[Ref.~\cite{GKS23}]\lb{t:catalytic-asymptotic}
Consider a resource theory with a set of free operations $\bbO$.
Suppose that $\rho$ is convertible to $\rho'$ with a correlated catalyst with a vanishing error by a free operation in $\bbO$ and $\rho$ is distillable.
Then, $\rho$ is convertible to $\rho'$ by a marginal-asymptotic conversion with a rate arbitrarily close to 1 by a free operation in $\bbO$.
}

Below we briefly outline the proof of this theorem.
By assumption, $\rho$ is convertible to $\rho'$ with a correlated catalyst with state $c$ by a free operation in $\bbO$.
Since $\rho$ is distillable, there exists a finite number $\mu \in \bbN_+$ such that $\rho^{\otimes \mu}$ is convertible to the catalyst $c$ with an arbitrarily small error by an operation in $\bbO$.
Now we construct the desired marginal-asymptotic conversion protocol as follows.
Our protocol converts $N$ copies of $\rho$ to $N-\mu$ copies of $\rho'$.
We first use $\mu$ copies of $\rho$ and obtain $c$.
Then, using the catalyst $c$ repeatedly, we convert $N-\mu$ copies of $\rho$ into $N-\mu$ copies of $\rho'$, which satisfies the condition of marginal-asymptotic conversions.
The conversion rate is $1-\frac \mu N$, which can be arbitrarily close to 1 by taking $N$ sufficiently large.

We note that although the obtained catalyst $c$ has a small error, the above protocol indeed works.
A crucial fact is that thanks to the monotonicity of the trace norm the error in the catalyst never increases through a repeated use of this catalyst.

\bigskip

These two propositions demonstrate that a correlated-catalytic conversion and a marginal-asymptotic conversion with rate 1 are two faces of the same coin, as long as the state is distillable.
It is worth recalling another connection between asymptotic conversions and catalytic conversions that an exact asymptotic conversion (i.e., the final state is ${\rho'}^{\otimes \lfloor rN\rfloor}$ exactly) with rate 1 implies an exact uncorrelated-catalytic conversion~\cite{Dua05}.

We here notice that a correlated-catalytic conversion can be obtained even from a marginal-asymptotic conversion, and the existence of an asymptotic conversion is not necessary.
This distinction is particularly important in the case that an asymptotic conversion with rate 1 does not exist while a marginal-asymptotic conversion with rate 1 exists.
This indeed happens in the resource theory of (unspeakable) coherence, which we will see in \sref{cov-review}.

\bigskip

We finally comment on the size of a catalyst and the number of copies for desired conversions.
As seen from above, these two are closely related.
We consider a resource theory with a set of free operations $\bbO$ and employ the same symbols $V$ as in the proof of \pref{Wil} (i.e., $V$ is a set of states convertible from $\rho$ with a catalyst with a vanishing error).
In the conversion from $\rho$ to $\rho'$, the necessary size of a catalyst and the necessary number of copies tend to be large if $\rho'$ is close to the boundary of $V$.
The speed of the increase of the number of copies (also the size of a catalyst) is evaluated as the convergence speed in the asymptotic theory.
In contrast, the necessary size of a catalyst and the necessary number of copies might be not so large when $\rho'$ is not close to the boundary of $V$.
The small size regime is usually not studied, since the behavior is considered to be non-universal and highly dependent on the details of the setup.

We present an example in the case of Gibbs-preserving operations (GPO) with a correlated-catalyst with a vanishing error (\pref{sGPO-c} in \sref{GPO-quantum}), where the necessary and sufficient condition of state conversion from $\rho$ to $\rho'$ is given by the order of the free energy: $F(\rho)\geq F(\rho')$.
The following example is first presented in Ref.~\cite{SS21}.
Consider a two-level system spanned by $\{ \ket{0}, \ket{1}\}$, and set 
\balign{
\rho&=\frac{3}{200} \ket{0}\bra{0}+\frac{197}{200} \ket{1}\bra{1} \\
\rho'&=\ket{+}\bra{+}
}
with $\ket{+}:=\frac{1}{\sqrt{2}}(\ket{0}+\ket{1})$, $\beta=1$, $E_0=0$ and $E_1=\ln 3$, implying the Gibbs state as $\rhoG=\frac34\ket{0}\bra{0}+\frac14\ket{1}\bra{1}$.
Free energies of $\rho$ and $\rho'$ are $F(\rho)= 1.291\cdots$ and $F(\rho')=0.836\cdots$, respectively.
We also set the upper bound of the error and the correlation strength as $\ep=0.01$ and $\delta=0.06$, respectively.
Note that since this final state sits at the boundary of the set of achievable states from $\rho$ and the error $\ep=0.01$ is small, the required size of a catalyst and the required number of copies might be large.

In this setting, a Gibbs-preserving operation without catalyst cannot convert $\rho$ to $\rho'$ with error less than $\ep=0.01$ (see \cite{SS21}).
On the other hand, there exists a Gibbs-preserving operation converting 8 copies of $\rho$ to 8 copies of $\rho'$ with error less than $\ep=0.01$.
More precisely, there exists a Gibbs-preserving operation converting 8 copies of $\rho$ to 
\eq{
\Xi:=(0.994\cdots)\ket{+}\bra{+}^{\otimes 8}+(0.005\cdots) \[ \frac{1}{1-\frac{25}{4^8}}\( \rhoG^{\otimes 8} -\frac{25}{4^8}\ket{+}\bra{+}^{\otimes 8}\) \] ,
}
which satisfies $d_1(\Xi, \ket{+}\bra{+}^{\otimes 8})<\ep$.
Using \pref{asymptotic-catalytic}, we find that a catalyst with 10 spins (7 copies of the system and a label system with $8=2^3$ states) enables us to implement a correlated-catalytic conversion from $\rho$ to $\rho'$.

\subsection{Gibbs-preserving operations}\lb{s:GPO-review}

We here review results on Gibbs-preserving operations (GPO).
We first describe GPO in the classical regime, and then treat GPO in the fully quantum regime.

\subsubsection{GPO in classical regime}\lb{s:GPO-classical}

We first consider GPO in the classical regime, where the state is given by a probability vector $\bsp$, and the set of free operations $\bbO$ is a set of stochastic matrices $T\in \bbO$ ($\bsp \to T\bsp$) which keeps the Gibbs distribution invariant.
Before stating the results, we first introduce several notions and quantities which are used to state these results.

\bdf{[Lorenz curve]
For a pair of two probability distribution $(\bsp, \bsq)$, we define $\bsp^*$ and $\bsq^*$ as the simultaneous rearrangement of $\bsp$ and $\bsq$ such that $p^*_1/q^*_1\leq p^*_2/q^*_2\leq \cdots$.
We define the Lorenz curve of  $(\bsp, \bsq)$ as a curve connecting $(0,0)$, $(p_1^*, q_i^*)$, $(p_1^*+p_2^*, q_1^*+q_2^*)$, $\ldots$, $(1,1)$.
}

\bdf{[extended free energy]
We define the extended free energy as
\eq{
F_\alpha(\bsp):=\frac1\beta (S_\alpha(\bsp ||\pG)-\ln Z).
}
Here, $Z:=\sum_i e^{-\beta E_i}$ is the partition function, and we defined the R\'{e}nyi $\alpha$-divergence as
\eq{
S_\alpha(\bsp ||\bsq):=\frac{{\rm sgn}(\alpha)}{\alpha-1}\ln (\sum_i p_i^\alpha/q_i^{\alpha-1})
}
for $\alpha\neq 0,1,\pm\infty$, and those with $\alpha= 0,1,\pm\infty$ as their limits.
}

$S_1(\bsp||\bsq)$ is the conventional relative entropy, and thus the extended free energy with $\alpha=1$ is the conventional nonequilibrium free energy, which we also denote by $F(\bsp)$.

\bpro{[State convertibility of single-shot GPO]\lb{t:Blackwell}
A probability distribution $\bsp$ is convertible to $\bsp'$ through a GPO exactly if and only if the Lorenz curve of $(\bsp, \pG)$ lies always above that of $(\bsp', \pG)$.
}

Note that \pref{Blackwell} is a special case of Blackwell's theorem, which is proven in a variety of ways~\cite{Bla51, Bla53, Cam, Vei, RM, RSS, MOAbook, AU, Cre, Joe, LS, Dah, Shi20}.

\bpro{[State convertibility of GPO with an uncorrelated catalyst~\cite{Bra15}]\lb{t:second-laws}
A probability distribution $\bsp$ is convertible to $\bsp'$ through a GPO with an uncorrelated catalyst with a vanishing error if and only if 
\eq{
F_\alpha (\bsp)\geq F_\alpha(\bsp')
}
holds for all $-\infty< \alpha<\infty$.
}

\bpro{[State convertibility of GPO with a correlated catalyst~\cite{Mul18}]\lb{t:c-second-law}
A probability distribution $\bsp$ is convertible to $\bsp'$ through a GPO with a correlated catalyst with a vanishing error if and only if 
\eq{
F(\bsp)\geq F(\bsp').
}
}

Notably, we recover the second law of thermodynamics in the framework with a correlated catalyst.
In other words, all other constraints in \pref{Blackwell} and \pref{second-laws} are removed by the existence of a catalyst and negligibly small correlation, and only the second law of thermodynamics robustly remains.

In the proof of \pref{second-laws}, we construct an elaborated catalyst with which the condition in \pref{Blackwell} (more precisely, its equivalence with a different expression) is satisfied.
In the proof of \pref{c-second-law},  we construct an elaborated catalyst with which the condition in \pref{second-laws} is satisfied.
In other words, in this proof approach, \pref{c-second-law} is built on \pref{second-laws}, and \pref{second-laws} is built on \pref{Blackwell}.

\subsubsection{GPO in quantum regime}\lb{s:GPO-quantum}

One may expect to extend these results to the fully quantum regime simply by quantizing these results.
In particular, it is conjectured that the second law of thermodynamics is also the unique condition of state convertibility in the fully quantum regime~\cite{WGE17, MS19}.
However, unfortunately, we cannot simply quantize these proofs, since the necessary and sufficient condition of state conversion without catalyst~\cite{Bus15} (the quantum counterpart of \pref{Blackwell}) is highly complicated and not useful to treat.
In particular, the quantum counterpart of \pref{second-laws} has been elusive, which prevent to show the quantum counterpart of \pref{c-second-law} in this approach.

Fortunately, a completely different approach from above resolves this conjecture~\cite{SS21}.
In the new approach, we first construct an asymptotic conversion with rate 1 and then construct the desired correlated-catalytic conversion by applying \pref{asymptotic-catalytic}.

\bdf{[Nonequilibrium free energy]
Consider a system $S$ with its Hamiltonian $H$.
Its Gibbs state at inverse temperature $\beta$ is denoted by $\rhoG:=e^{-\beta H}/Z$ with $Z:=\Tr[e^{-\beta H}]$.
Using them, the nonequilibrium free energy of $\rho$ is defined as $F(\rho):=S(\rho||\rhoG)=\Tr[\rho\ln \rho-\rho\ln \rhoG]$.
}

\bpro{[Asymptotic conversion by GPO (used in \cite{Fai19, SS21}): \pref{GPO} in the main text]\lb{t:sGPO}
Consider two states $\rho$ and $\rho'$ in $S$ with $F(\rho)\geq F(\rho')$.
Then, for any $\delta_1>0$ there exists a large $N$ and a GPO $\Lambda: S^{\otimes N}\to S^{\otimes N}$ such that $\abs{\Lambda(\rho^{\otimes N})-\rho'^{\otimes N}}<\delta_1$.
}

\bpro{[Correlated-catalytic conversion by GPO~\cite{SS21}]\lb{t:sGPO-c}
Consider two states $\rho$ and $\rho'$ in $S$ with $F(\rho)\geq F(\rho')$.
Then, for any $\ep>0$ and $\delta>0$ there exist a catalyst $C$, its state $c$, and a GPO $\Lambda\in \bbO: SC\to SC$ such that 
\eq{
\Lambda(\rho\otimes c)=\tau
}
with $\Tr_S[\tau]=c$, $\abs{\Tr_C[\tau]-\rho'}_1<\ep$, and $\abs{\tau-\Tr_C[\tau]\otimes c}_1<\delta$.
}

As clearly seen from \pref{sGPO-c}, the second law of thermodynamics is also recovered in the fully quantum regime with the help of the correlated catalyst.

\subsection{Covariant operations}\lb{s:cov-review}

We here review results on covariant operations.
This subject has been investigated in the context of the resource theory of (unspeakable) coherence, also known as the resource theory of asymmetry.

From the perspective of physics, the resource theory of (unspeakable) coherence aims to clarify possible/impossible operations under the law of energy conservation, which is an unavoidable ubiquitous constraint in our world.
The covariant condition is a direct consequence of the law of energy conservation, and we cannot create coherence among energy eigenstates from scratch without any help.
We note that the covariant operation with energy conservation is mathematically formulated as an $U(1)$-symmetric operation.

We here briefly comment on a technical point in the resource theory of (unspeakable) coherence.
In this resource theory, it is frequently assumed that all the energy level spacings in the system are relatively rational (i.e., for any level spacings $ij$ and $kl$, we have $\Delta E_{ij}/\Delta E_{kl}\in \bbQ$)~\cite{GS08, Mar20, Mar22, Abe14}.
In other words, there is a fixed real number $\Di$ such that for any level spacing $ij$ there exists an integer $a_{ij}\in \bbN$ such that $\Delta E_{ij}=a_{ij}\Di$.
From a mathematical aspect, this assumption is equivalent to the existence of a time period, stating that there exists a finite $\tau$ such that $e^{-iH\tau}\rho e^{iH\tau}=\rho$ for any $\rho$.
Of course, this is an idealization and actual matters in physics usually have irrational level spacings due to inevitable impurities.
However, we notice that actual systems inevitably have ambiguities in the value of eigenenergies due to the measurement errors, and owing to the denseness of rational numbers any actual system has corresponding a system with rational level spacings within this errors.
From this background, it is considered to be justifiable similarly to other frequently appearing idealizations in physics, such as a shift-invariant clean lattice.

\bigskip

Previous studies have found highly contrastive results on the power of covariant operations.
On the one hand, it has been suggested that the barrier of the coherence is very severe and the covariant operation has very weak power.
On the other hand, recent studies have revealed that in slightly stretched setups the barrier of the coherence almost vanishes and the covariant operation realizes almost all operations.
In the following, we shall see these two contrastive faces of covariant operations in the catalytic framework and the asymptotic framework one by one.

\subsubsection{Catalytic framework}

Previous studies on conversions without catalyst and with an uncorrelated catalyst suggest that the barrier of the coherence is very severe and the covariant operation has very weak power~\cite{Mar-thesis, MS13, MS14}.
In addition, even with the help of a correlated catalyst, it is proven that an incoherent initial state never results in a coherent state, which is called {\it no broadcasting theorem} of coherence and asymmetry:

\bpro{[No broadcasting theorem~\cite{LM19, MS19})]\lb{t:no-broadcast}
Suppose that $\rho$ is incoherent and $\rho$ is converted to $\rho'$ through a covariant operation with a correlated catalyst.
Then, $\rho'$ is also incoherent.
}

From these results, coherence manipulation among different energy eigenstates under the law of energy conservation appears to be a severe task.
This implies that we will encounter many barriers (various constraints) in quantum operations under the law of energy conservation.

On the other hand, it is also revealed that covariant operations in the correlated-catalytic framework admit almost all conversions are implementable, and the initial incoherent states belongs to its exceptional case.

\bpro{[Correlated-catalytic conversion by covariant operations~\cite{ST23, KGS23}]\lb{t:coherence-trivial}
Suppose that any coherent mode in $\rho'$ can be expressed by a sum of integer multiples of coherent modes in $\rho$.
Then, a covariant operation with a correlated catalyst convert $\rho$ to $\rho'$ with a vanishing error.
}

This result suggests that the existence of coherence under the law of energy conservation provides almost no barrier to our manipulations, and covariant operations can realize almost all state conversions.
The no-broadcasting theorem treats its exceptional case, incoherent initial states.

We prove \pref{coherence-trivial} by applying \pref{asymptotic-catalytic}, stating the connection between correlated-catalytic conversions and marginal-asymptotic conversions, to \pref{coherence-marginal} shown in the next subsection, which claims the existence of the covariant marginal-asymptotic conversions with conversion rate 1.

\subsubsection{Asymptotic framework}

The power of asymptotic conversions by covariant operations shows highly contrastive results between asymptotic conversions and marginal-asymptotic conversions:

\bpro{[Asymptotic conversion by covariant operations~\cite{Mar20}]\lb{t:distill-impossible}
Suppose that the initial state $\rho$ is  a full-rank state and  the final state $\rho'$ is a pure coherent state $\rho'=\ket{\phi}\bra{\psi}$.
Then, the conversion rate from $\rho$ to $\rho'$ by an asymptotic conversion is zero.
In other words, there exists $\ep>0$ such that for any positive $r>0$ there exists $N$ such that for any $N'\geq N$ all covariant operations $\Lambda: S^{\otimes N'}\to S^{\otimes \lfloor rN'\rfloor}$ map $\rho^{\otimes N}$ to states far from ${\rho'}^{\otimes \lfloor rN'\rfloor}$ more than distance $\ep$: $\abs{\Lambda(\rho^{\otimes N})-{\rho'}^{\otimes \lfloor rN'\rfloor}}_1>\ep$.
}

\bpro{[Marginal-asymptotic conversion by covariant operations~\cite{ST23, KGS23}]\lb{t:coherence-marginal}
Suppose that any coherent mode in the final state $\rho'$ can be expressed by a sum of integer multiples of coherent modes in the initial state $\rho$.
Then, the conversion rate from $\rho$ to $\rho'$ by a marginal-asymptotic conversion is unbounded.
In other words, for any $\ep>0$ and any positive $r>0$, there exists a sufficiently large $N$ and a covariant operation $\Lambda: S^{\otimes N}\to S^{\otimes \lfloor rN\rfloor}$ such that $\Lambda(\rho^{\otimes N})=\Xi$ satisfying $\abs{\Tr_{\bcs i}[\Xi]-\rho'}_1<\ep$ for all $i\in \{1, 2,\ldots ,\lfloor rN\rfloor\}$.
}

In the asymptotic conversion, if the initial state is a full-rank state and the final state is a pure coherence state, we cannot achieve any finite conversion rate even if the rate is very small (e.g., $\ep=0.001$).
In contrast, in the marginal-asymptotic conversion, if the initial state has (maybe small but) finite coherence, we can obtain any final state with any large rate.
The latter protocol allows us a counterintuitive conversion that a few almost incoherent initial states are converted into many maximally coherent states.


\bigskip

We here present another result partially filling these two gap: asymptotic conversion with any sublinear rate.
Note that the definition of the error in the following theorem is the same as the asymptotic conversion.

\bpro{[Asymptotic conversion by covariant operations with a sublinear rate~\cite{Mar20}]\lb{t:sublinear}
Consider a system such that all energy level spacings are integer multiples of a fixed value $\Di$.
Assume that the shortest period of the initial state $\rho$ is $2\pi/\Di$.
Then, for any $\rho'$, any function $M(N)$ satisfying $\lim_{N\to \infty}\frac{M(N)}{N}=0$, and any $\ep>0$, there exists a large $N'$ such that for any $N\geq N'$ there exists a covariant operation $\Lambda: S^{\otimes N}\to S^{\otimes M(N)}$ satisfying $\abs{\Lambda(\rho^{\otimes N})-{\rho'}^{\otimes M(N)}}_1<\ep$.
}


The key idea to prove this theorem is the protocol of covariant time estimation.
In the covariant time estimation, the Hamiltonian $H$ and state $\rho$ is fixed, and state $\rho(t)=e^{-iHt}\rho e^{iHt}$ is given.
Our task is to estimate $t$ from many copies of $\rho(t)$ by using only covariant operations.
The following theorem confirms that there exists a covariant time (phase) estimation protocol whose variance decays as the inverse of the number of copies.

\bpro{[Covariant time estimation (used in \cite{Mar20}): \pref{cov} in the main text]\lb{t:scov}
Consider a system such that all energy level spacings are integer multiples of a fixed value $\Di$.
Assume that  the shortest period of the initial state $\rho$ is $2\pi/\Di$.
Then, there exists a time estimation protocol $S^{\otimes m}\to \bbR$ with probability distribution $P(\test|\kappa)$ for $\kappa$ on $S^{\otimes m}$ such that (i) covariant: $P(\test|e^{-iH^{\otimes m}\tau}\kappa e^{iH^{\otimes m}\tau})=P(\test+\tau|\kappa)$ for any $\kappa$ and $\tau$, and (ii) the variance of $P(\test|\rho^{\otimes m})$ decays as $O(1/m)$.
}

This estimation protocol is covariant in the following sense that if the phase of the input state $\kappa$ is shifted by $\tau$ as $\kappa\to e^{-iH^{\otimes m}\tau}\kappa e^{iH^{\otimes m}\tau}$, then the output estimator is also shifted as $\test\to \test+\tau$.

Without loss of generality, we suppose that the average of $\test$ for $\rho$ is zero (we set the origin of the phase as supposition).
The decaying speed $O(1/m)$ is the same as the central limit theorem, and thus this decaying speed is an expected behavior.

\medskip

Now we construct an asymptotic conversion protocol claimed in \pref{sublinear}.
In this sublinear rate conversion protocol, we first estimate $\test$ by using the protocol shown in \pref{scov} by using state $\rho^{\otimes N}$, and then prepare $e^{-iH^{\otimes M}\test}(\rho')^{\otimes M}e^{iH^{\otimes M}\test}$.
Notice that the distance between $\rho'$ and $e^{-iH\test}\rho'e^{iH\test}$ decays as $O(1/N)$, and $\abs{(\rho')^{\otimes M}-e^{-iH^{\otimes M}\test}(\rho')^{\otimes M}e^{iH^{\otimes M}\test}}_1$ is bounded from above by $M\abs{\rho'-e^{-iH\test}\rho'e^{iH\test}}_1$ due to the subadditivity of the trace norm distance with respect to the tensor-product.
This observation suggests that the output error behaves as $\frac MN$, which converges to zero by assumption, and thus the output state is arbitrarily close to $(\rho')^{\otimes M}$, which completes the proof of \pref{sublinear}.


\section{Proof of \lref{lemma}}
\subsection{Statement}
We consider a system whose energy level spacings are relatively rational, which is a frequently employed assumption in the resource theory of (unspeakable) coherence~\cite{Mar20, Mar22, GS08, Abe14} (see also \sref{cov-review} for its further explanation).
This assumption is equivalent to the requirement that all the level spacings are integer multiples of a fixed value $\Di$.
We assume that $\Di$ is the maximum of such values, i.e., we cannot replace $\Di$ by $n\Di$ with an integer $n\geq 2$.
In other words,  any state $\rho$ has a finite period $\tau=2\pi /\Di$, i.e., $e^{-iH\tau}\rho e^{iH\tau}=\rho$, and there exists a state whose shortest period is $2\pi /\Di$.
Here, we normalize the Planck constant to unity.

In this section, we consider a marginal-asymptotic conversion by a covariant Gibbs-preserving operation (CGPO).
We first define the CGPO, which is an operation satisfying both the Gibbs-preserving condition and the covariant condition.

\bdf{[Covariant Gibbs-preserving operation (CGPO)]
A map $\calE: S\to S$ is called a covariant Gibbs-preserving operation (CGPO) if 
\bi{
\item \ul{Gibbs-preserving}: Let $\rhoG:=e^{-\beta H}/Z$ be a Gibbs state with a given inverse temperature $\beta$. Then, we have
\eq{
\calE(\rhoG)=\rhoG.
}

\item \ul{Covariant}: For any $t$, we have
\eq{
\calE(e^{-iHt}\rho e^{iHt})=e^{-iHt}\calE(\rho) e^{iHt}.
}
}
}

Similarly to the case of GPO, if the operation maps the system $S$ to a different system $S'$, then the condition of Gibbs-preserving is replaced by $\calE(\rhoG)=\rhopG$.

As seen in \sref{GPO-review} and \sref{cov-review}, both Gibbs-preserving operations (GPO) and covariant operations are independently studied in depth (see also Refs.~\cite{Los-review, Shi25}).
The CGPO is an operation class of their intersection, which is also called {\it enhanced thermal operation} in Ref.~\cite{Cwi15} and studied in several papers~\cite{Gou18, DDH21, WT24}.
The CGPO, or enhanced thermal operation, is an operation class close to thermal operation while it is mathematically easy to treat.

\bigskip


Defining the free energy $F(\rho):=S(\rho||\rhoG)$ with the quantum relative entropy $S(\sigma||\eta):=\Tr[\sigma \ln \sigma -\sigma \ln \eta]$~\cite{Haybook}, we are ready to state a key lemma, which is the same as \lref{lemma} in the main text.

\blm{[\lref{lemma} in the main text]\lb{t:slemma}
Consider two states $\rho$ and $\rho'$ in $S$ whose energy level spacings are integer multiples of $\Di$.
We assume that the shortest period of $\rho$ is $2\pi/\Di$ (i.e., all modes are coherent) and $F(\rho)\geq F(\rho')$.
Then, for any $\delta >0$ there exists a marginal-asymptotic conversion from $\rho$ to $\rho'$ through CGPO with conversion rate $1-\delta$ with a vanishing error.
}

In the following, we first construct the desired conversion protocol and then demonstrate that this protocol is indeed covariant, Gibbs-preserving, and realizing the desired state conversion with arbitrarily small error.

\subsection{Construction of the protocol}\lb{s:construct-marginal}

We construct a covariant Gibbs-preserving protocol converting $N$ copies of $\rho$ into $(1-\delta)N$ copies of $\rho'$ for sufficiently large $N$.
Our key idea is the combination of \pref{scov} (asymptotic covariant conversion with sublinear rate) and \pref{sGPO} (asymptotic Gibbs-preserving conversion).
To implement CGPO, we sandwich the asymptotic Gibbs-preserving conversion protocol by the covariant time estimation and time shift protocols.
However, to realize a desired conversion with arbitrarily small error, we need additional cares.

We decompose $N$ copies of $\rho$ into $\nu:=(1-\delta)\sqrt{N}$ sets of $\sqrt{N}$ copies and two $\delta N/2$ copies, which we call $A$ part, $B_1$ part, and $B_2$ part respectively.
The $\nu$ sets in $A$ part are labeled as $A_1,\ldots , A_\nu$.
In addition, we introduce a time propagator by $t$ for a state $R$ on $S^{\otimes n}$ defined as
\eq{
\calT_t(R):=e^{-iH^{\otimes n}t} R e^{iH^{\otimes n}t}.
}
The number $n$ is properly chosen depending on state $R$ such that if $R$ is a state of $S^{\otimes \alpha}$, then we set $n=\alpha$.

Our marginal-asymptotic conversion protocol $\calN: S^{\otimes N}\to S^{\otimes (1-\delta)N}$ consists of two steps:
\be{
\item We apply the covariant time estimation protocol shown in \pref{scov} to $B_1$ part and $B_2$ part (two $\delta N/2$ copies of $\rho$) and obtain $\test^1$ and $\test^2$.
The variances of $\test^1$ and $\test^2$ decay as $O(1/\delta N)$.
\item We apply the map
\eqa{
\calM_{\test^1,\test^2}( \rho^{\otimes \nu\sqrt{N}}):=\calT_{\test^2}\circ \calT_{-\test^1} (\rho^{\otimes \nu \sqrt{N}})
}{protocol}
on $A$ part.
Here, $\Lambda$ is the Gibbs preserving map converting $\rho^{\sqrt{N}}$ to $\rho'^{\sqrt{N}}$, whose existence is shown in \pref{sGPO} with a fixed error $\delta_1$.
}
Hence, the entire map is given by
\balign{
\calN(\rho^{\otimes N})=&\int  d\test^1 d\test^2 P(\test^1|\rho^{\otimes \delta N/2})P(\test^2|\rho^{\otimes \delta N/2})\calM_{\test^1,\test^2}( \rho^{\otimes(1-\delta)N}) \nt \\
=&\( \int d\test^2 P(\test^2|\rho^{\otimes \delta N/2}) \calT_{\test^2} \) \circ \Lmd^{\otimes \nu} \circ \( \int  d\test^1 P(\test^1|\rho^{\otimes \delta N/2})\calT_{-\test^1} (\rho^{\otimes(1-\delta)N})\) \lb{entire-N}
}

Before demonstrating desired properties, we here explain the idea behind this construction.
In each set $A_i$, the map $\calT_{\test^2}\circ \Lmd \circ \calT_{-\test^1} (\rho^{\otimes \sqrt{N}})$ works in parallel.
Since we employ the same $\test^1$ and $\test^2$ in all sets, the resulting states of different sets, in $A_i$ and in $A_j$, are correlated.
However, fortunately, since the correlation among sets does not matter in the definition of the error of marginal-asymptotic conversions, it suffices to examine whether the conversion in a single set realizes the desired conversion.

Thus, it suffices to focus on a single set $A_i$, and $B_1$ and $B_2$ parts.
We set the size of $A_i$ such that it is sublinear with respect to the size of $B_1$ and $B_2$, which enables us to adopt a similar argument to \pref{sublinear} (sublinear rate conversion with a vanishing error).
To compensate the smallness of each set, we prepare sufficiently many sets enough to make the total number of copies in $A$ part overwhelming compared to that of $B_1$ and $B_2$.

We finally comment on the structure of the map in a single set.
In the examination of the covariant condition, we apply a phase shift $\calT_{\tau}$ on the initial state.
Two phase shift operations, $\calT_{-\test^1}$ and $\calT_{\test^2}$, play the role to cancel the additional phase at the first stage and to recover the additional phase at the last stage, respectively.
Thanks to this trick, the input states of the GPO, $\Lambda^{\otimes \nu}$, are always the same regardless of the addition of the phase.


\subsection{Demonstrating the desired properties}
\subsubsection{Covariant property}
We first demonstrate that the constructed protocol $\calN$ satisfies the covariant condition, i.e., for any state $\Sigma$ on $S^{\otimes N}$ and any real number $\tau$, 
\eqa{
\calN(\calT_{\tau}(\Sigma))=\calT_{\tau}(\calN(\Sigma))
}{cov-N}
is satisfied.

Denote by $\Sigma_A$, $\Sigma_{B_1}$, and $\Sigma_{B_2}$ the reduced state of $\Sigma$ to $A$, $B_1$, and $B_2$, respectively.
Let
\eqa{
\kappa:=\int  d\test^1 P(\test^1|\Sigma_{B_1})\calT_{-\test^1}(\Sigma_A)
}{def-kappa}
be the state on which $\Lmd^{\otimes \nu}$ acts (see \eref{entire-N}).
The final state is written as
\eqa{
\eta:=\calN(\Sigma)=\int  d\test^2 P(\test^2|\Sigma_{B_2})\calT_{\test^2}\circ \Lmd^{\otimes \nu}(\kappa).
}{def-eta}

Now, in order to confirm the covariant condition, we replace the input state $\Sigma$ by $\calT_\tau(\Sigma)=e^{-iH^{\otimes N}\tau}\Sigma e^{iH^{\otimes N}\tau}$.
Note that when $\Sigma_{B_i}$ is replaced by $\Sigma_{B_i}\to \calT_\tau(\Sigma_{B_i})$, the stochastic variable $\test^i$ is shifted as $\test^i\to \test^i+\tau$ with the same probability distribution.
Since the new estimator $\test'^1$ follows $P(\test'^1|\calT_\tau(\Sigma_{B_1}))=P(\test'^1-\tau|\Sigma_{B_1})$, when our initial state is $\calT_\tau(\Sigma_{A})$, the input state for $\Lambda^{\otimes \nu}$ denoted by $\kappa_\tau$ is computed as
\balign{
\kappa_\tau=&\int  d\test'^1 P(\test'^1|\calT_\tau(\Sigma_{B_1}))\calT_{-\test'^1}(\calT_\tau(\Sigma_{A})) 
=\int  d\test^1 P(\test^1|\Sigma_{B_1})\calT_{-\test^1-\tau}( \calT_\tau(\Sigma_A)) 
=\kappa.
}
This means that the additional time shift $\tau$ by $\calT_\tau$ on $\Sigma_A$ is completely canceled by the shift of the time estimator $\calT_{-\test'^1}=\calT_{-\test^1-\tau}$ from $\calT_{-\test^1}$, and the input state for $\Lambda^{\otimes \nu}$ is always the same state $\kappa$ regardless of the time shift $\tau$.

Following a similar argument, the final state denoted by $\eta_\tau$ is computed as
\eq{
\eta_\tau=\int  d\test'^2 P(\test'^2|\calT_{\tau}(\Sigma_{B_2}))\calT_{\test'^2}\circ \Lmd^{\otimes \nu}(\kappa_\tau)=\int  d\test'^2 P(\test^2|\calT_{\tau}(\Sigma_{B_2}))\calT_{\test^2+\tau}\circ \Lmd^{\otimes \nu}(\kappa)=\calT_{\tau}(\eta).
}
This means that the additional time shift $\tau$ is put to the final state through $\calT_{\test'^2}$, and the covariant condition \eqref{cov-N} is indeed satisfied.

\bigskip

We note that the independence of two estimators $\test^1$ and $\test^2$ is necessary in the above calculation.
In fact, if we use the same estimator twice (i.e., employing $\calT_{\test}\circ \Lmd^{\otimes \nu} \circ \calT_{-\test}$), the first shift $\calT_{-\test}$ and the second shift $\calT_{\test}$ are correlated, and this conversion is no longer covariant in general.

\subsubsection{Gibbs-preserving property}

We next demonstrate that this protocol satisfies the Gibbs-preserving condition, that is,
\eqa{
\calN(\rhoG^{\otimes N})=\rhoG^{\otimes (1-\delta)N}.
}{GPO-N}

This property is confirmed by solely examining the map $\calM$ in \eref{protocol}.
Since $\rhoG$ is diagonal with respect to the energy eigenbasis, $\rhoG$ is invariant under time propagation $\calT$.
Hence, for any $\test^1$ and $\test^2$, the Gibbs-preserving property in each set $A_i$ is demonstrated as follows:
\eq{
\calT_{\test^2}\circ \Lambda\circ \calT_{-\test^1} (\rhoG^{\otimes \sqrt{N}})=\calT_{\test^2}\circ \Lambda(\rhoG^{\otimes \sqrt{N}})=\calT_{\test^2}(\rhoG^{\otimes \sqrt{N}}) =\rhoG^{\otimes \sqrt{N}}.
}
In the second equality, we used the fact that $\Lambda$ is a GPO.
Our final state is the $(1-\delta)\sqrt{N}$ copies of the above state, which is equivalent to the right-hand side of \eref{GPO-N}.

\subsubsection{Conversion from $\rho$ to $\rho'$ with arbitrary accuracy}\lb{s:CGPO-accuracy}

We finally demonstrate that this protocol realizes the desired marginal-asymptotic conversion with arbitrarily small error.
In each set $A_i$, we apply three maps, $\calT_{-\test^1}$, $\Lambda^{\otimes \nu}$, and $\calT_{\test^2}$, successively.
The reference conversion (the ideal situation) of these three maps is
\eq{
\rho^{\otimes \nu\sqrt{N}}\to \rho^{\otimes \nu\sqrt{N}}\to {\rho'}^{\otimes \nu\sqrt{N}}\to {\rho'}^{\otimes \nu\sqrt{N}},
}
meaning that both $\calT_{-\test^1}$ and $\calT_{\test^2}$ are close to identity maps and $\Lambda$ converts $\rho$ to $\rho'$ with high accuracy.
Of course, in our actual protocol, all of these three maps have errors.
Since the time estimation is performed on a finite number (precisely, $\delta N/2$) of copies, estimators $\test^1$ and $\test^2$ have errors around $\test^1=\test^2=0$, which causes $\calT_{-\test^1}$ and $\calT_{\test^2}$ deviating from identity maps.
In addition, the prepared Gibbs-preserving map $\Lambda$ has an inherent error, which is supposed to be less than $\delta_1$ by construction.
We need to take the contributions from these three errors and bound the error in the final state.

To demonstrate the accuracy of our conversion, we first introduce a distance of quantum states different from the trace norm (see Ref.~\cite{Haybook} for details).
We here employ the Bures distance 
\eq{
b(\rho, \sigma):=\sqrt{1-F(\rho, \sigma)},
}
which is defined by using the fidelity 
\eq{
F(\rho, \sigma):=\Tr[\sqrt{\sqrt{\rho}\sigma\sqrt{\rho}}].
}
It is known that the Bures distance satisfies the axiom of distance.
The Bures distance behaves as $b^2(\rho, \calT_t(\rho))\simeq \frac{J}{8}t^2$ for small $t$, where $J$ is the SLD Fisher information.
This leads to the fact that the square of the Bures distance $b^2(\rho, \calT_t(\rho))$ is bounded from above by the square of $t$ with a proper coefficient, that is, we have 
\eq{
b^2(\rho, \calT_t(\rho))\leq Ct^2
}
with some constant $C$.
In addition, the fidelity for tensor product states behaves as 
\eq{
F(\rho^{\otimes m}, \sigma^{\otimes m})=(F(\rho, \sigma))^m.
}
Combining these relations, in a small $t$ regime we find 
\balign{
b(\rho^{\otimes m}, \calT_t(\rho^{\otimes m}))=b(\rho^{\otimes m}, \calT_t(\rho)^{\otimes m})=\sqrt{1-[F(\rho, \calT_t(\rho))]^m}=\sqrt{1-[1-b^2(\rho, \calT_t(\rho))]^m}\leq &\sqrt{1-(1-Ct^2)^m} \nt \\
=& O(\sqrt{m}t). \lb{Bures-shift-error}
}

Now we show that this protocol indeed realizes the desired marginal-asymptotic conversion with an arbitrarily small error.
Since the marginal-asymptotic conversion cares about only the error in a single copy, the correlation among different sets caused by the use of the common estimators $\test^1$ and $\test^2$ is irrelevant in the evaluation of the conversion error, and it suffices to treat a single set $A_i$.
\pref{scov} shows that the error of $\test^1$ and $\test^2$ (i.e., $\abs{\test^1}$ and $\abs{\test^2}$) are of order $O(1/\sqrt{\delta N})$.
Combining this evaluation and \eref{Bures-shift-error}, we find that the state $\calT_{-\test^1} (\rho^{\otimes \sqrt{N}})$ is close to $\rho^{\otimes \sqrt{N}}$ as
\eqa{
b(\rho^{\otimes \sqrt{N}}, \calT_{-\test^1} (\rho^{\otimes \sqrt{N}}))=O\(\sqrt{\sqrt{N}} \frac{1}{\sqrt{\delta N}}\) =O\( \frac{1}{\sqrt{\delta\sqrt{N}}}\)
}{Bures-error}
in terms of the Bures distance.
For the same reason, $\calT_{\test^2}\circ \Lambda(\kappa)$ is close to $\Lambda(\kappa)$ with error $O(1/\sqrt{\delta\sqrt{N}})$ in terms of the Bures distance.

Denote the input state to $\Lambda$ in each set $A_i$ by
\eq{
\zeta:=\int  d\test^1 P(\test^1|\rho^{\delta N/2})\calT_{-\test^1}(\rho^{\otimes \sqrt{N}}).
}
We notice that the final state in each set $A_i$ is expressed as
\eq{
\Tr_{\bcs A_i}[\eta]=\int  d\test^2 P(\test^2|\rho^{\delta N/2})\calT_{\test^2}\circ \Lmd(\zeta).
}
Then, the error in each set $A_i$ from $\rho'^{\otimes \sqrt{N}}$ in terms of the trace norm is evaluated as
\balign{
\abs{\Tr_{\bcs A_i}[\eta]-\rho'^{\otimes \sqrt{N}}}_1\leq&  \abs{\Tr_{\bcs A_i}[\eta]-\Lmd(\zeta)}_1+\abs{\Lmd(\rho^{\otimes \sqrt{N}})-\rho'^{\otimes \sqrt{N}}}_1+\abs{\Lmd(\zeta)-\Lmd(\rho^{\otimes \sqrt{N}})}_1 \nt \\
\leq&  \abs{\Tr_{\bcs A_i}[\eta]-\Lmd(\zeta)}_1+\abs{\Lmd(\rho^{\otimes \sqrt{N}})-\rho'^{\otimes \sqrt{N}}}_1+\abs{\zeta-\rho^{\otimes \sqrt{N}}}_1 \nt \\
\leq&  2\sqrt{2}b(\Tr_{\bcs A_i}[\eta], \Lmd(\zeta))+\abs{\Lmd(\rho^{\otimes \sqrt{N}})-\rho'^{\otimes \sqrt{N}}}_1+ 2\sqrt{2}b(\zeta, \rho^{\otimes \sqrt{N}}) \nt \\
<&O\( \frac{1}{\sqrt{\delta \sqrt{N}}}\)+\delta_1+O\( \frac{1}{\sqrt{\delta \sqrt{N}}}\) . \lb{CGPO-error-final}
}
In the second line, we used the monotonicity of the trace norm.
In the third line, we used a simple relation $\abs{\rho-\sigma}_1\leq 2\sqrt{2}b(\rho, \sigma)$~\cite{Haybook}.
In the fourth line, we applied \eref{Bures-error} and the assumed relation $\abs{\Lambda(\rho^{\otimes \sqrt{N}})-\rho'^{\otimes \sqrt{N}}}_1<\delta_1$.
The three sources of the error, the fluctuation from the phase shift $\calT_{-\test^1}$, the error in the Gibbs-preserving map $\Lambda$, and the fluctuation from the phase shift $\calT_{\test^2}$, correspond to the first, second, and third terms in the right-hand side, respectively.
Since for any $\delta>0$, we can take arbitrarily small $\delta_1$ and arbitrarily large $N$, the above bound implies an arbitrarily small error.

\section{Proof of \tref{main}}\lb{s:proof-main}
We here show that \lref{lemma} (\lref{slemma}) implies \tref{main}.
Below, we recast \tref{main}:

\bthm{[\tref{main} in the main text]\lb{t:smain}
Consider two states $\rho$ and $\rho'$ in $S$ whose energy level spacings are integer multiples of $\Di$.
We assume that the shortest period of $\rho$ is $2\pi/\Di$ (i.e., all modes are coherent).
Then, $\rho$ is convertible to $\rho'$ through CGPO with a correlated-catalyst with a vanishing error if and only if $F(\rho)\geq F(\rho')$.

In addition, if $F(\rho)>F(\rho')$ and $\rho'$ is full-rank, then this conversion is exact.
}


We first demonstrate the former part of \tref{smain}.
Recall that \pref{sGPO-c} shows that the necessary and sufficient condition of state conversion $\rho\to \rho'$ with a correlated-catalyst by GPO is $F(\rho)\geq F(\rho')$.
Since CGPO is a subset of GPO, it suffices to prove the sufficient part of \tref{smain}.
Fortunately, we know that the sufficient part of \tref{smain} with an arbitrarily small error can be obtained by applying \pref{asymptotic-catalytic} (asymptotic conversion with rate 1 implies correlated-catalytic conversion) to \lref{slemma}.

\bigskip

We next demonstrate the latter part of \tref{smain};  the existence of an exact conversion if $F(\rho)>F(\rho')$ and $\rho'$ is full-rank.
This fact is shown by applying \pref{Wil} (see \sref{catalytic-asymptotic}).
We set $S_m$ as a set of states convertible from $\rho$ exactly through a CGPO with a correlated catalyst whose Hilbert space has a dimension less than or equal to $m$.
Then, $V=(\lim_{m\to \infty}S_m)^{\rm c}$ is a set of states achievable from $\rho$ by a correlated-catalytic conversion with an arbitrarily small error.
We notice that the boundary of $V$ consists of (i) low-rank states, and (ii) states $\rho'$ with $F(\rho)=F(\rho')$.
Hence, a full-rank $\rho'$ with $F(\rho)>F(\rho')$ is an interior point of $V$, and thus \pref{Wil} suggests $\rho'\in S_m$ for some $m$.
This means that $\rho'$ is obtained from $\rho$ with a correlated-catalyst with dimension $m$ exactly, which completes the proof of the latter part of \tref{smain}.

\bigskip

In summary, we explain how to construct the desired catalyst for given $\rho$ and $\rho'$ satisfying the supposition of \tref{smain} and acceptable error $\ep>0$.
The catalyst can be prepared by the following steps:
\be{
\item Let $N_{\rm GPO}$ be the minimum number such that there exists a GPO $\Lambda$ satisfying 
\eq{
\abs{\Lambda(\rho^{\otimes N_{\rm GPO}})-{\rho'}^{\otimes N_{\rm GPO}}}<\ep/6.
}
Let $N_{\rm est}$ be the minimum number such that for any quantum state $\zeta$ on $S$ we have 
\eq{
\int d\test P(\test|\rho^{\otimes \ep N_{\rm est}/4})\abs{\zeta^{\otimes \sqrt{N_{\rm est}}}-\calT_{\test}(\zeta^{\otimes \sqrt{N_{\rm est}}} )}_1<\ep/6.
}
We set $N=\max(N_{\rm GPO}, N_{\rm est})$.
\item We prepare a CGPO converting $\rho^{\otimes N}$ to ${\rho'}^{\otimes (1-\ep/2)N}$ in the sense of marginal-asymptotic conversion with error $\ep$, which is constructed in \sref{construct-marginal} with $\delta=\frac\ep2$.
Precisely, we first consume two $N\ep/4$ copies of $\rho$ and estimate $\test^1$ and $\test^2$ by a covariant time estimation.
Then, using these two estimators we apply $\calM$ in \eref{protocol} to $\rho^{\otimes \sqrt{N}}$.
Since we have $(1-\ep/2)\sqrt{N}$ sets of $\rho^{\otimes \sqrt{N}}$, we apply $\calM$ by $(1-\ep/2)\sqrt{N}$ times.
\item We construct the catalyst in the line with \pref{asymptotic-catalytic} for the above marginal-asymptotic conversion.
Precisely, our catalyst is on $S^{\otimes N-1}\otimes R$ with a label system $R$, and set the state of the catalyst as \eref{catalyst-construction}.
}

\section{Proof of \tref{gen}}\lb{s:gen}
We shall prove a generalization of \tref{smain} (\tref{main} in the main text) to general resource theories, which is presented as \tref{gen} in the main text.
To present the precise statement of the theorem and its proof, we first define the admittance of the phase estimation and the phase shift by a resource theory.
We employ the same symbols $P(\test|\kappa)$ and $\calT_{\test}$ as \tref{main} with modifying this general setting.

\bdf{[Admittance of the phase estimation and the phase shift]\lb{t:phase-gen}
Consider system $S$ whose energy level spacings are integer multiples of $\Di$.
We say that a resource theory with a set of free operation $\bbO$ admits the phase estimation and the phase shift if for any given coherent reference state $\rho\in S$ whose shortest period is $2\pi/\Di$ there exists a phase estimation protocol and the phase shift protocol in $\bbO$ as follows:
\bi{
\item \ul{Phase estimation}: We can implement a time estimation protocol $S^{\otimes m}\to \bbR$ with probability distribution $P(\test|\kappa)$ for $\kappa$ on $S^{\otimes m}$ by a free operation in $\bbO$ such that (i) covariant: $P(\test|e^{-iH^{\otimes m}\tau}\kappa e^{iH^{\otimes m}\tau})=P(\test+\tau|\kappa)$ for any $\kappa$ and $\tau$, and (ii) the variance of $P(\test|\rho^{\otimes m})$ decays as $O(1/m)$ for any $\rho$ in $S$ whose shortest period is $2\pi/\Di$.
\item \ul{Phase shift}: We can implement the map $\calT_t(\rho)=e^{-iHt}\rho e^{iHt}$ by a free operation in $\bbO$.
}
}

We put the admittance of the phase estimation and the phase shift as an assumption, since we need to implement the phase estimation protocol and the phase shift protocol (shown in \pref{scov} and below) in this resource theory.


Now we generalize \tref{smain} to general resource theories with a set of free operations $\bbO$:

\bthm{[\tref{gen} in the main text]\lb{t:sgen}
Consider a resource theory whose set of free operations is $\bbO$ and system $S$ whose energy level spacings are integer multiples of $\Di$.
Suppose that $\rho$ is distillable, and has shortest period $2\pi/\Delta$.
In addition, the phase estimation and the phase shift can be performed in $\bbO$ in the sense of \dref{phase-gen}.

Then, if $\rho$ is convertible to $\rho'$ with a correlated catalyst with a vanishing error in a resource theory by $\bbO$, then $\rho$ is also convertible to $\rho'$ with a correlated catalyst with a vanishing error by $\bbO\cap {\rm Cov}$.
}

This theorem claims that as long as the suppositions (coherent and distillable initial state, the admittance of phase estimation and phase shift) are satisfied, the state convertibility by $\bbO$ and that by $\bbO\cap {\rm Cov}$ are the same in the correlated-catalytic framework.
This means a striking fact that the addition of the covariant condition provides no change on the resource theory in general.
For example, since the resource theory of entanglement with LOCC admits the phase estimation and phase shift, if the initial state is coherent and distillable, the resource theory with LOCC and that with LOCC$\cap {\rm Cov}$ have the same power of state conversions.
This theorem answers why we can usually ignore the existence of the law of energy conservation, which inevitably accompanies the covariant condition, while obtained results on resource theories are physically meaningful.
The answer by \tref{sgen} is that the state convertibility without the covariant condition and that with the covariant condition are the same, and thus we can safely investigate the latter setup without taking the covariant condition into account.

\bigskip

The construction of this conversion protocol is essentially the same as that for \tref{smain} (and \lref{slemma}).
As seen in \sref{proof-main}, it suffices to construct a marginal-asymptotic conversion protocol by a free operation in $\bbO$ from $\rho$ to $\rho'$ with a conversion rate arbitrarily close to 1.
To this end, we employ \pref{catalytic-asymptotic}, stating that a marginal-asymptotic conversion for distillable states with rate 1 implies a correlated-catalytic conversion.
Let $\Lambda$ be the marginal-asymptotic map converting $\rho^{\otimes \gamma}$ to $\rho'^{\otimes \gamma}$, whose existence is confirmed by \pref{catalytic-asymptotic}.
Then, constructing the same protocol as \eref{protocol} we obtain the desired marginal-asymptotic map in $\bbO \cap{\rm Cov}$.
The idea of this construction is summarized as
\begin{quote}
(correlated-catalytic conversion by $\bbO$) \\
 $\xrightarrow{\pref{catalytic-asymptotic}}$ (marginal asymptotic conversion with rate 1 by $\bbO$) \\
 $\xrightarrow{protocol \ as \ \eref{protocol}}$ (marginal asymptotic conversion with rate 1 by $\bbO \cap{\rm Cov}$) \\
$\xrightarrow{\pref{asymptotic-catalytic}}$ (correlated-catalytic conversion by $\bbO\cap{\rm Cov}$)
\end{quote}

For completeness, we describe the constructed marginal-asymptotic conversion protocol by $\bbO \cap{\rm Cov}$ below:
\be{
\item We apply the covariant time estimation protocol shown in \pref{scov} on $B_1$ part and $B_2$ part (two $\delta N/2$ copies of $\rho$) and obtain $\test^1$ and $\test^2$.
The variances of $\test^1$ and $\test^2$ decay as $O(1/\delta N)$.
\item We apply the map
\eqa{
\calM( \rho^{\otimes \nu\sqrt{N}}):=\calT_{\test^2}\circ \Lmd^{\otimes \nu} \circ \calT_{-\test^1} (\rho^{\otimes \nu \sqrt{N}})
}{protocol}
on $A$ part.
Here, $\Lambda$ is the free operation in $\bbO$ converting $\rho^{\sqrt{N}}$ to $\rho'^{\sqrt{N}}$, whose existence is confirmed by \pref{catalytic-asymptotic}.
}
Applying \pref{asymptotic-catalytic} to this protocol, we obtain the desired correlated-catalytic protocol converting $\rho$ to $\rho'$ by an operation in $\bbO \cap{\rm Cov}$.


\end{document}